\documentclass[11pt]{article}

\usepackage[T1]{fontenc}
\usepackage[utf8]{inputenc}
\usepackage[margin=1in]{geometry}
\usepackage{microtype}
\usepackage{authblk}

\usepackage{amsmath,amssymb,amsfonts,amsthm,mathtools}

\usepackage{graphicx}
\graphicspath{{images/}}
\usepackage{booktabs}
\usepackage{multirow}
\usepackage{array}
\usepackage{tabularx}
\usepackage{longtable}
\usepackage{rotating}
\usepackage{float}

\usepackage{xcolor}
\usepackage{enumitem}

\usepackage[numbers,sort&compress]{natbib}
\usepackage{url}
\usepackage[hidelinks]{hyperref}

\theoremstyle{definition}

\theoremstyle{remark}

\usepackage{algorithm}
\usepackage{algpseudocode}

\providecommand{\keywords}[1]{%
  \par\vspace{0.5em}\noindent\textbf{Keywords: }#1\par
}

\title{Longitudinal Random Forests for Sparse and Irregular Response Trajectories}

\author[1]{Yangsheng Wang}
\author[2]{Xiaotian Dai}
\author[3]{Haoda Fu}
\author[1]{Guifang Fu\thanks{Corresponding author: gfu@binghamton.edu}}

\affil[1]{%
Department of Mathematics and Statistics, Binghamton University,
Binghamton, New York 13902, USA}

\affil[2]{%
Department of Mathematics, Illinois State University,
Normal, Illinois 61790, USA}

\affil[3]{%
Department of Biostatistics, The University of North Carolina at Chapel Hill,
Chapel Hill, North Carolina 27599, USA}

\date{}

\begin{document}

\maketitle

\begin{abstract}
Longitudinal studies often collect data at sparse, irregular, and unequally spaced time points. Such heterogeneity is often driven by subject-specific covariates, yet existing methods have been restricted to a scalar endpoint value, completely neglecting the underlying response trajectories. We propose a novel Longitudinal Random Forest (LRF) framework that leverages tree-based ensemble machine learning with adaptive node-wise longitudinal trajectory estimation. The LRF framework makes five methodological contributions: it captures each subject's individual response trajectory while simultaneously accommodating within-node correlation, between-node heterogeneity, and nonlinear and interactive covariate effects; it introduces a novel trajectory-based splitting criterion that maximizes trajectory separation while incorporating a size-weighted penalty; it provides two variants, Principal Analysis by Conditional Expectation (LRF--PACE) and adaptive linear mixed-effects models (LRF--adaptiveLMM), which employ nonparametric and semiparametric node-wise smoothers, respectively, while learning covariate effects in a data-driven manner; it provides a comprehensive interpretation of covariates using both the classical trajectory-based permutation variable importance measure (PVIM) and a newly proposed finite-way interaction frequency count; and it not only predicts entire trajectories for new subjects but also forecasts future trajectories for existing subjects. Extensive simulation studies demonstrate that LRF achieves superior performance over several competing methods, even under severe sparsity. The practical significance of the LRF framework lies in its ability to address five important clinical questions. To the best of our knowledge, it is the first method to model underlying longitudinal response trajectories for characterizing treatment durability in diabetes clinical trials, representing a revolutionary advance over traditional endpoint efficacy analysis. LRF has important implications for personalized medicine through individualized model structures, individualized response trajectories, and subject-specific covariate effects.
\end{abstract}

\keywords{Random Forests | Longitudinal Data Analysis | Sparse and Irregular Observations | Clinical Trials | Nonparametric/Semiparametric Statistics}


\section{Introduction}

Diabetes mellitus is one of the most prevalent chronic diseases worldwide and is associated with severe complications, reduced life expectancy, and substantial public health and economic costs \citep{sun2022idf}. Long-term glycemic control is commonly assessed by glycated hemoglobin (HbA1c), which reflects the average blood glucose level over approximately the preceding 90 days of the measurement time \citep{eyth2025hemoglobin}. Repeated HbA1c measurements characterize how glycemic levels change over time after treatment initiation and reflect the durability of treatment effects \citep{sherwani2016significance, eyth2025hemoglobin, Kim2019}. This work is motivated by a de-identified longitudinal diabetes clinical trial dataset derived from an insulin-initiation study (INSULIN trial), a large multinational study comparing glycemic responses to two treatments in adults with type 2 diabetes (T2D) \citep{buse2009durable, buse2011durable}. T2D is a heterogeneous disease, with individuals exhibiting markedly different responses to the same treatment \citep{mccarthy2017personalised, pearson2019multifaceted, QIU2025102367}. Such heterogeneity may arise from variations in subjects’ baseline metabolic, anthropometric, hemodynamic, glucose profiles, disease duration, as well as genetic factors, lifestyle, etc.

In many clinical studies, however, such heterogeneity is mainly evaluated using endpoint efficacy, which is a scalar response defined as the response value at the final scheduled measurement minus its baseline value. This approach has been commonly applied in existing T2D studies, including large phase 3 clinical trials such as SURPASS, SUSTAIN, and AWARD \citep{Frias2021tirzepatide, Sorli2017SUSTAIN3, Dungan2014AWARD6}. Two widely used statistical approaches for analyzing endpoints were analysis of covariance (ANCOVA) and the mixed model for repeated measures (MMRM). ANCOVA estimated treatment effects on endpoint efficacy while adjusting for covariates \citep{vickers2001analysing}. MMRM accounted for repeated measurements through a marginal covariance structure; however, it was primarily used to estimate treatment effects on the population mean response at the final scheduled visit, rather than to model subject-specific response trajectories \citep{mallinckrodt2008recommendations}. Prior analyses of the same INSULIN trial dataset have primarily focused on personalized treatment recommendations based on HbA1c endpoint efficacy. For example, \citet{wang2018learning} proposed two learning frameworks for determining personalized treatments based on a scalar HbA1c endpoint efficacy response value. One framework fit regression models to estimate subject-specific treatment contrasts under a prespecified risk constraint, while the other employed an outcome-weighted learning approach to estimate treatment strategies through inverse probability weighted classification. Similarly, \citet{doubleday2018algorithm} developed a univariate random forest–based approach that employed an inverse probability weighted (IPW) function to determine candidate splits and established an individualized treatment strategy based on the higher estimated HbA1c endpoint efficacy value.

In practice, the final scheduled measurement is often unavailable because subjects discontinue treatment before completing all follow-up visits, so it has to be imputed to estimate the endpoint efficacy. Existing approaches included single-imputation methods such as last observation carried forward (LOCF), baseline or worst observation carried forward (BOCF/WOCF), as well as multiple-imputation procedures that generated plausible endpoint values from distributions estimated on the repeatedly observed measurements \citep{horton2007much, he2010missing, yan2025comparison}. Consequently, endpoint efficacy was not uniquely defined and may vary across studies depending on study-specific analytic choices \citep{sassi-sayadi2026regulatory}. In addition, subjects with the same endpoint value may exhibit markedly different HbA1c trajectories following treatment initiation \citep{Tee2023trajectory, Luo2018hba1c}. This highlights a fundamental limitation of scalar endpoint efficacy because it cannot capture the durability of a treatment.

Long-term durability reflects not only the magnitude of glycemic reduction but also the rate and overall trajectory of HbA1c change over time. This perspective motivates several important clinical questions: 1) How to enable precision medicine by assigning treatments based on each patient's baseline characteristics; 2) How to characterize the durability of HbA1c response following treatment initiation; 3) Which patients are most likely to benefit from a given treatment; 4) How to predict a patient's HbA1c response durability throughout the treatment period before treatment is actually initiated; and 5) how to forecast future HbA1c response values after the treatment initiation period.

The well-established longitudinal data analysis approaches in the existing literature were designed to capture underlying trajectories from repeated measurements while accounting for both within-subject correlation and between-subject heterogeneity \citep{diggle2002analysis, fitzmaurice2011applied, caruana2015longitudinal, Tabarraei2024}. However, they did not yet fully address several important challenges arising in real-world clinical applications: First, longitudinal data are often sparsely and irregularly observed, with different subjects measured at different time points and contributing varying numbers of observations, which complicates estimation of the underlying mean and covariance functions \citep{fan2007semiparametric, chen2015irregular, delporte2024analysing}. Second, repeated measurements within a subject are inherently correlated, and the correlation structure may vary over time and differ among subjects, making it difficult to model subject-specific longitudinal trends accurately \citep{lu2010covariance, verbeke2014multivariate, savieri2025limmcov}. Third, when incorporating covariates into longitudinal response trajectories, complex nonlinear and interactive covariate effects complicate model specification.

Generalized estimating equations (GEE) and linear mixed-effects models (LMM) were two of the most widely used traditional parametric statistical approaches for longitudinal data analysis \citep{liang1986longitudinal, laird1982random}. GEE extended generalized linear models to longitudinal data by specifying a marginal mean structure together with a marginal working correlation structure to accommodate repeated measurements. However, it did not model subject-specific trajectories and hence did not consider between-subject heterogeneity \citep{Mancl2001, wang2016covariance, DieuBriganti2025GEE}. In comparison, LMM introduced subject-specific random effects to account for within-subject correlation and to model individual longitudinal trajectories \citep{VerbekeMolenberghs2009, vanderHorn2024}. However, both GEE and LMM assumed linear trajectory trends and linear covariate effects. These assumptions may be restrictive when HbA1c trajectories are highly dynamic or when covariate effects are nonlinear and interactive \citep{Tee2023trajectory, Handley2025BMJOpen, McCoy2023Trajectories, Aniley2019SemiParametric}.

Traditional nonparametric statistical approaches estimated each subject's longitudinal response as a smooth function of time without imposing any prespecified forms. Methods such as Principal Analysis by Conditional Expectation (PACE) were particularly well suited for sparse and irregularly observed longitudinal data. PACE first estimated smooth population-level mean and covariance functions and then obtained subject-specific functional principal component (FPC) scores through conditional expectation given each subject’s observed measurements \citep{yao2005functional}. However, PACE was simply a method for estimating trajectories alone without involving any covariates. PACE-based functional regression first extracted subject-specific FPC scores and then connected these estimated FPC scores with covariates using a regression model \citep{yao2005functional}. This two-step approach separated trajectory feature extraction from covariate modeling, so information from covariates was not incorporated when estimating subject-specific trajectories.

In this article, we propose a novel Longitudinal Random Forest (LRF) framework that extends tree-based ensemble learning to longitudinal data by leveraging adaptive longitudinal estimation approaches and machine learning techniques. The proposed framework makes five methodological contributions: First, this framework captures each subject's individual response trajectory, while simultaneously accommodating within-node correlation, between-node heterogeneity, nonlinear and interactive covariate effects, and sparse and irregularly observed repeated measurements. Second, to separate heterogeneous trajectories, we propose a novel trajectory-based splitting criterion that selects candidate splits by maximizing the integrated squared difference between the estimated representative trajectories of the resulting child nodes, while incorporating a size-weighted penalty to prevent unstable splits driven by random noise. Third, we develop two flexible variants, LRF--PACE and LRF--adaptiveLMM, which differ in their node-wise longitudinal smoothers while both allowing covariate effects to be learned in a data-driven manner. Specifically, LRF--PACE is a fully nonparametric approach that flexibly models dynamic response trajectories within every node of the tree structure, whereas LRF--adaptiveLMM is a semiparametric approach that embeds an LMM at each node of the tree structure, with node-specific fixed-effect model determined adaptively by the variables selected for splitting along the path from the root to each node. 

Fourth, inspired by univariate random forests, we design a trajectory-based permutation variable importance measure (PVIM) to rank covariates according to the change in out-of-bag prediction error, measured over the repeated measures along the response trajectory, before and after permuting each covariate individually. In addition, we propose a new frequency-based interaction detection strategy that identifies covariate interactions by tracking how frequently each multi-way covariate combination jointly appears along the same tree paths across all trees fitted over all cross-validation replications. Existing permutation-based approaches for interaction detection typically require jointly permuting multiple covariates as a whole unit, leading to $\binom{p}{k}$ number of combinations when assessing $k$-way interactions. When $p$ is large, this number becomes enormous even for 2-way interactions. Moreover, the procedure must be repeated separately for each $k$-way interaction. In contrast, the proposed frequency-based interaction detection strategy is computationally efficient and enables simultaneous detection of any finite multi-way interactions in a single procedure. Fifth, the proposed LRF framework predicts the response trajectory over the entire observed time period for new subjects using only their covariates. Moreover, it forecasts future unobserved response values beyond the observed time window for existing subjects using their covariates and observed longitudinal response history.

Several tree-based methods for longitudinal data have been developed in the literature, but they differ fundamentally from the proposed LRF framework in how longitudinal trajectories and tree structures are incorporated. These existing approaches fall into two broad categories: semiparametric and nonparametric methods. The semiparametric category combined mixed-effects models with recursive partitioning. For example, Stochastic Random Effects Expectation Maximization Forest (SREEMF) embedded random forests into a mixed-effects model by replacing the fixed-effects term with random forests to estimate a nonlinear mean structure after adjusting for random effects \citep{hajjem2011mixed, hajjem2014mixed, sela2012reem, sela2021reemtree, capitaine2020longiturf, capitaine2021rf}. Longitudinal Classification and Regression Tree (LongCART) partitioned subjects through parameter instability tests and estimated longitudinal trends only at the terminal nodes using a parsimonious linear mixed-effects model containing only an intercept and a time effect term \citep{kundu2019regtree}. By comparison, LRF--adaptiveLMM adaptively determines a linear mixed-effects model at each node, simultaneously leveraging the tree structure, allowing different covariate effects and corresponding linear mixed-effects models across different nodes. Consequently, subjects assigned to the same node share the same fixed-effect structure while generating individualized trajectories through their subject-specific random effects. Neither SREEMF nor LongCART explicitly incorporated covariates into node-specific mixed-effects models. Among the nonparametric category, SplineForest (SPLF) first estimated individual trajectories using smoothing splines and subsequently applied multivariate random forests to the resulting spline coefficients \citep{YuLambert1999, NeufeldHeggeseth2025}. Since smoothing splines were generally designed for densely observed data, the resulting trajectory estimation may lose accuracy under severely sparse and irregularly observed schedules. Furthermore, SPLF separated trajectory estimation from covariate learning through a two-step procedure. In contrast, the proposed LRF--PACE framework jointly performs covariate partitioning and longitudinal trajectory estimation within a unified framework through a trajectory-based splitting criterion.

We evaluate the proposed LRF models through two simulation settings under various levels of sparsity. For variable selection, we compare the LRF models with SREEMF and SPLF, as only these two methods provide variable importance measures. For prediction, we consider SREEMF, SPLF, LongCART, and classical LMM and GEE as competing approaches. The simulation results show that the LRF models accurately identify all the true covariates and achieve high prediction and forecast accuracy, even under high missingness and complex nonlinear interaction effects. We further apply the proposed LRF framework to the INSULIN trial to demonstrate its practical utility through a real data analysis. The practical significance of the proposed LRF framework lies in its ability to address the five important clinical questions described above. In addition, to the best of our knowledge, this is the first method to model longitudinal response trajectories for characterizing treatment durability in diabetes clinical trials, representing a revolutionary advance over traditional endpoint efficacy analysis.

\section{Motivating Dataset} \label{sec:data}
As the motivating data, the INSULIN trial \citep{buse2009durable, buse2011durable}, is a 24-week, randomized, open-label, multinational study comparing the efficacy and durability of two insulin treatments in adults with T2D inadequately controlled on oral antidiabetic agents. The trial enrolled $2,091$ insulin-naive participants from $242$ centers across 11 countries. Participants were randomly assigned to one of two different treatments and followed for up to 24 weeks, with repeated HbA1c measurements collected during the initiation phase. Among the enrolled subjects, $1,520$ subjects with complete baseline covariate information were retained for analysis. A total of 19 baseline covariates were measured, encompassing metabolic, anthropometric, hemodynamic, and glucose profile measures, etc. The metabolic category includes four core biomarkers: baseline A1C, fasting insulin, fasting blood glucose (FG), and adiponectin. The anthropometric category comprises body weight, height, and body mass index (BMI). The hemodynamic category includes heart rate, diastolic blood pressure, and systolic blood pressure. To capture same-day glycemic variation, seven self-monitored blood glucose (SMBG) covariates are incorporated: morning fasting glucose, 2-hour post-breakfast glucose, noon fasting glucose, 2-hour post-lunch glucose, evening fasting glucose, 2-hour post-dinner glucose, and 3 a.m. fasting glucose. In addition, diabetes duration and a categorical treatment assignment (Therapy 1 vs. Therapy 2) are included to account for disease history and treatment choice.

The HbA1c measurements in the INSULIN trial data exhibit a highly irregular missingness pattern over time. Measurements are nearly complete during the initial few weeks. All subjects are measured at Week~1, after which subjects follow one of the two prespecified visit schedules: 97\% are measured at Weeks~3 and~5, whereas the remaining 3\% are measured at Weeks~2 and~4. After Week 5, observation rate decreases substantially, with only 25\%--55\% of subjects observed at visits between Weeks 6 and 14. Measurements are almost empty between Weeks 15 and 22, with 0\% observations recorded at these seven weeks except Week 20 that has fewer than 1.2\% of subjects observed. Observation rates rise again to approximately 35\% at Weeks 23--24, where the final scheduled measurement ends. Overall, the near-empty interval between Weeks 15 and 22, spanning approximately ${1}/{3}$ of the entire interval, presents substantial challenges for reliable trajectory estimation.

\begin{figure*}[t]
\centering
\includegraphics[width=0.7\linewidth]{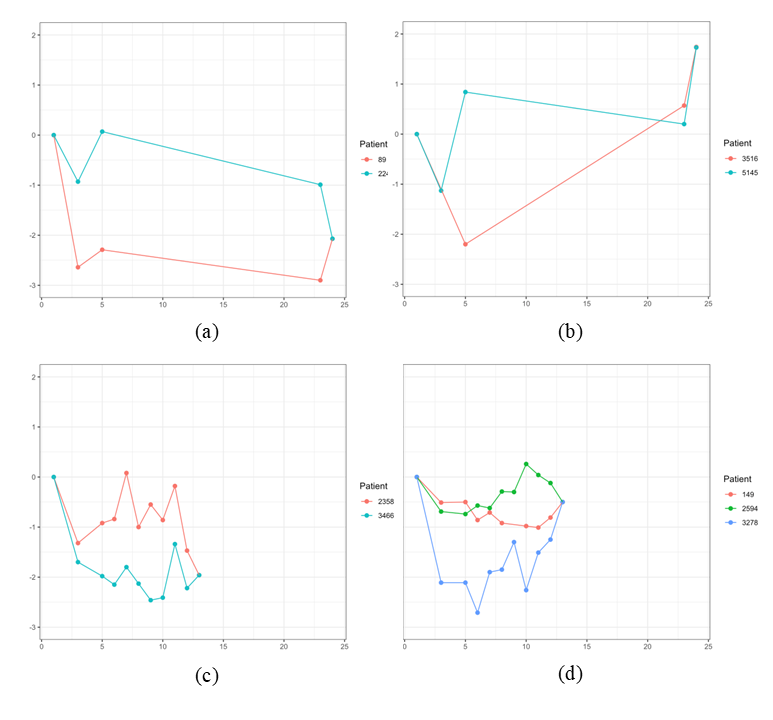}
\caption{The original discretely and repeatedly measured HbA1c values for nine representative subjects, illustrating the heterogeneous characteristics of the longitudinal INSULIN trial. Panels show (a) decreased HbA1c value at Week 24, (b) increased HbA1c value at Week 24, (c) decreasing HbA1c value at Week 14, and (d) similar HbA1c value at Week 14.}
\label{fig:hb_overall}
\end{figure*}

\begin{figure*}[t]
\centering
\includegraphics[width=0.8\textwidth]{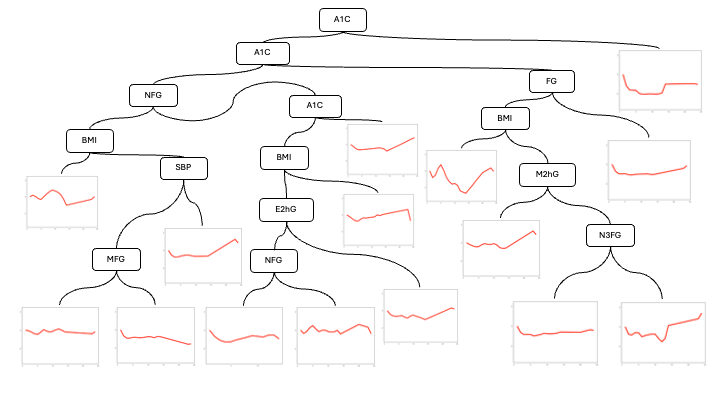}
\caption{A representative tree obtained by fitting LRF--PACE to the INSULIN trial data. The covariates appearing in the tree are selected through the proposed trajectory-based splitting optimization procedure. The red curves represent the node-specific estimated HbA1c trajectories, which are modeled nonparametrically as functions of covariates while accommodating nonlinear and interactive covariate effects as well as heterogeneous response trajectories.}
\label{fig:durable_tree}

\vspace{0.3em}

\begin{minipage}{\linewidth}
\footnotesize
\raggedright
\textit{Abbreviations:}
Metabolic covariates: A1C = baseline glycated hemoglobin (HbA1c); FG = fasting glucose. SMBG measurements: MFG = morning fasting glucose; M2hG = morning 2-hour postprandial glucose; NFG = noon fasting glucose; E2hG = evening 2-hour postprandial glucose; N3FG = 3 a.m.\ fasting glucose. Anthropometric covariate: BMI = body mass index. Hemodynamic covariate: SBP = systolic blood pressure.
\end{minipage}
\end{figure*}

Throughout this paper, HbA1c response is defined as each post-baseline HbA1c measurement minus the subject-specific baseline value. Under this definition, all trajectories begin at a zero value. Figure \ref{fig:hb_overall} presents the HbA1c measurements for nine representative subjects and illustrates several important characteristics of the longitudinal data. First, the total numbers and detailed schedules of sampling differ across subjects. Panels (a) and (b) represent subjects with only 5 out of the 24 scheduled measurements, observed primarily during the early study phase (up to Week 5), followed by a completely empty period between Weeks 6 and 22, and regain the last two observations at Weeks 23 and 24 at the end of the trial. In comparison, Panels (c) and (d) show subjects with relatively dense follow-up during Weeks 1--14, contributing more than 10 out of the 24 scheduled measurements during this period, but with no further observations during the remainder of the study period. Second, Figure \ref{fig:hb_overall} illustrates heterogeneity both in endpoint values and in temporal trajectories. Specifically, the last observed HbA1c response value may decrease (Panels (a) and (c)), increase (Panel (b)), or remain nearly unchanged (Panel (d)) compared with its baseline value. Notably, the same endpoint values may arise from fundamentally different longitudinal trajectories, underscoring the importance of modeling the full trajectory rather than relying solely on a scalar endpoint. For example, in Panel (c), one subject exhibits an initial decrease followed by a rebound and subsequent decline, whereas the other shows a more gradual overall decrease.

To further illustrate the motivation behind the proposed LRF framework for capturing heterogeneous longitudinal response trajectories as a function of baseline covariates, Figure~\ref{fig:durable_tree} shows a representative tree obtained by fitting LRF--PACE to the INSULIN trial data. Subjects assigned to the same terminal node exhibit homogeneous response trajectories, summarized by the red representative curve within each node. The resulting tree structure reveals complex nonlinear and interactive covariate effects that partition heterogeneous longitudinal response trajectories into distinct terminal nodes. This process provides valuable insights for personalized medicine.

\section{Methodology}\label{sec:methodology}

We consider longitudinal data of the form
\[
\{(t_{ij}, Y_{ij}) : i = 1,\dots, N; ~j = 1,\dots, T_i\; \},
\]
where $Y_{ij}$ denotes the longitudinal response for subject $i$ discretely measured at time point $t_{ir}$, and $T_i$ is the subject-specific total number of time points. The measurement times $(t_{i1},\dots,t_{iT_i})$ vary in both spacing and timing across subjects, resulting in irregular, sparse, and unequal numbers of measurements. Let $\mathbf{X}_i=(X_{i1},\ldots,X_{ip})\in\mathbb{R}^{p}$ denote the covariate vector and $\mathbf{Y}_i=(Y_{i1},\ldots,Y_{iT_i})\in\mathbb{R}^{T_i}$ denote the longitudinal response vector for subject $i$, respectively. Our goal is to model and predict each subject’s response trajectory $\widehat{\mathbf{Y}}_i(t)$ conditional on $\mathbf{X}_i$, while capturing nonlinear and interactive covariate effects, and accommodating subject-specific heterogeneity, sparsity, and irregularity.

\subsection{Node-Specific Trajectory Estimation for Longitudinal Data}\label{sec:node_smoothing}

For a given node $\mathcal{I}$, let $\widehat{\mathbf{Y}}_{i,\mathcal{I}}(t)$ denote the smoothed response trajectory for subject $i$ estimated within that node. To accommodate different longitudinal modeling strategies, the LRF framework can incorporate either of two flexible node-wise smoothers: LRF--PACE or LRF--adaptiveLMM.

\subsubsection{LRF--PACE}

Within each node $\mathcal{I}$, subject-specific trajectories are estimated using the PACE framework, which provides nonparametric reconstruction of smooth functions for sparse and irregular longitudinal sampling. At each node, LRF--PACE treats the observed measurements as noisy realizations of an underlying smooth process and estimates node-specific mean and covariance functions via Gaussian kernel smoothing. Let $K_{h_{\mathcal I}}(\cdot)$ denote the Gaussian kernel function with bandwidth $h_{\mathcal I}$ that is selected by generalized cross-validation (GCV).

Specifically, for subject $i \in \mathcal{I}$ observed at times $t_{ij}$, the observed longitudinal response is modeled as
\[
Y_{ij}
=
\mu_{\mathcal{I}}(t_{ij})
+
\sum_{l=1}^{L}
\xi_{il,\mathcal{I}}
\phi_{l,\mathcal{I}}(t_{ij})
+
\varepsilon_{ij},
\qquad j=1,\ldots,T_i,
\]
where $\varepsilon_{ij}$ denote independent and identically distributed measurement errors with mean zero and variance $\sigma_{\mathcal{I}}^2$.
The node-specific mean function $\widehat{\mu}_{\mathcal{I}}(t)$ is obtained using a local linear smoother, which is defined as $\widehat{\mu}_{\mathcal{I}}(t)=\widehat{\alpha}_{\mathcal{I}}(t)$, where
\[
(\widehat{\alpha}_{\mathcal{I}}(t), \widehat{\gamma}_{\mathcal{I}}(t))
=
\arg\min_{\alpha,\gamma}
\sum_{i\in\mathcal{I}}\sum_{j=1}^{T_i}
K_{h_{\mathcal{I}}}(t_{ij}-t)
\bigl\{ Y_{ij} - \alpha - \gamma(t_{ij}-t) \bigr\}^2.
\]

The mean-centered observations within node $\mathcal{I}$ are defined as
\[
R_{ij,\mathcal{I}}
=
Y_{ij}-\widehat{\mu}_{\mathcal{I}}(t_{ij}),
\]
whose pairwise products are then smoothed through a two-dimensional kernel smoother to estimate the node-specific covariance surface
\[
\widehat{G}_{\mathcal{I}}(s,t)
=
\frac{
\sum_{i\in\mathcal{I}}\sum_{j\neq j'}
K_{h_{s,\mathcal{I}}}(t_{ij}-s)
K_{h_{t,\mathcal{I}}}(t_{ij'}-t)
R_{ij,\mathcal{I}}
R_{ij',\mathcal{I}}
}{
\sum_{i\in\mathcal{I}}\sum_{j\neq j'}
K_{h_{s,\mathcal{I}}}(t_{ij}-s)
K_{h_{t,\mathcal{I}}}(t_{ij'}-t)
}.
\]

Eigen-decomposition of $\widehat{G}_{\mathcal{I}}(s,t)$ yields estimated node-specific eigenvalues
$\{\widehat{\lambda}_{l,\mathcal{I}}\}$ and corresponding node-specific eigenfunctions
$\{\widehat{\phi}_{l,\mathcal{I}}(t)\}$, representing the dominant modes of temporal variation within node $\mathcal{I}$. The FPC score for subject $i$ and component $l$, is estimated by
\[
\widehat{\xi}_{il,\mathcal{I}}
=
\widehat{\lambda}_{l,\mathcal{I}}
\left(
\mathbf{Y}_i-\widehat{\boldsymbol{\mu}}_{i,\mathcal{I}}
\right)
\widehat{\Sigma}_{i,\mathcal{I}}^{-1}
\widehat{\boldsymbol{\phi}}_{il,\mathcal{I}}^{\top},
\qquad l=1,\ldots,L,
\]
where $L$ denotes the number of FPCs retained for estimation.
$\widehat{\boldsymbol{\mu}}_{i,\mathcal{I}}
=
\bigl(
\widehat{\mu}_{\mathcal{I}}(t_{i1}),\,
\ldots,\,
\widehat{\mu}_{\mathcal{I}}(t_{iT_i})
\bigr)$
and
$\widehat{\boldsymbol{\phi}}_{il,\mathcal{I}}
=
\bigl(
\widehat{\phi}_{l,\mathcal{I}}(t_{i1}),\,
\ldots,\,
\widehat{\phi}_{l,\mathcal{I}}(t_{iT_i})
\bigr)$
are obtained by evaluating the mean function and the $l$th eigenfunction at subject $i$'s observed time points, respectively.
The covariance matrix $\widehat{\Sigma}_{i,\mathcal{I}}$ has entries
\[
[\widehat{\Sigma}_{i,\mathcal{I}}]_{jj'}
=
\widehat{G}_{\mathcal{I}}(t_{ij},t_{ij'})
+
\widehat{\sigma}_{\mathcal{I}}^{2}\mathbf{1}\{j=j'\},
\]
where $\widehat{\sigma}_{\mathcal{I}}^{2}$ is the estimated variance of the measurement error $\varepsilon_{ij}$ within node $\mathcal{I}$, and $\mathbf{1}\{j=j'\}$ is an indicator function that equals 1 when $j=j'$ and 0 otherwise.

The reconstructed smooth trajectory generated by LRF-PACE for subject $i$ at node $\mathcal{I}$ is therefore
\begin{equation}
\widehat{\mathbf{Y}}_{i,\mathcal{I}}(t)
=
\widehat{\mu}_{\mathcal{I}}(t)
+
\sum_{l=1}^{L}
\widehat{\xi}_{il,\mathcal{I}}
\widehat{\phi}_{l,\mathcal{I}}(t).
\label{eq:pace_fitted_curve}
\end{equation}

\subsubsection{LRF--adaptiveLMM}

Within each node $\mathcal{I}$, subject-specific trajectories are estimated using a linear mixed-effects model, which provides a semiparametric representation of longitudinal trends. Specifically, the LMM model is fitted using a node-specific covariate set $\mathbf{X}_{i,\mathcal{I}}$, which contains the covariates selected along the branch path from the root to node $\mathcal{I}$ for subject $i$. Consequently, the fixed-effects structure in the LRF--adaptiveLMM is determined adaptively through the tree-growing process rather than prespecified in advance.

Then, for subject $i \in \mathcal{I}$ observed at times $t_{ij}$, the node-specific linear mixed-effects model is
\[
Y_{ij}
=
\beta_{0,\mathcal{I}}
+ \beta_{t,\mathcal{I}} t_{ij}
+ \mathbf{X}_{i,\mathcal{I}}^\top \boldsymbol{\beta}_{\mathcal{I}}
+ b_{i0,\mathcal{I}} + b_{i1,\mathcal{I}} t_{ij}
+ \varepsilon_{ij},
\qquad j=1,\dots,T_i,
\]
where $(b_{i0,\mathcal{I}}, b_{i1,\mathcal{I}})^\top$ are subject-specific random effects with covariance matrix $\mathbf{D}_{\mathcal{I}}$, and $\varepsilon_{ij}$ are independent and identically distributed measurement errors with mean zero and variance $\sigma_{\mathcal{I}}^2$.

Let $\mathbf{Z}_i = [\,\mathbf{1}_{T_i},\, \mathbf{t}_i\,]$ denote the random-effects design matrix, where $\mathbf{1}_{T_i}$ is a $T_i$-dimensional column vector of ones and $\mathbf{t}_i=(t_{i1},\ldots,t_{iT_i})^T$ is the vector of the observed time sampling for subject $i$. The subject-level covariance of the response vector $\mathbf{Y}_i$ at node $\mathcal{I}$ is
\[
\operatorname{Cov}(\mathbf{Y}_i^\top)
=
\mathbf{Z}_i\,\mathbf{D}_{\mathcal{I}}\,\mathbf{Z}_i^\top + \sigma^2_{\mathcal{I}} \mathbf{I}_{T_i},
\]
where $\mathbf{I}_{T_i}$ is the $T_i\times T_i$ identity matrix. This
covariance structure induces within-subject correlation through the random
intercept and random slope.

The fixed-effect parameters and covariance parameters are estimated by restricted maximum likelihood, and subject-specific random effects $(\widehat{b}_{i0,\mathcal{I}},\widehat{b}_{i1,\mathcal{I}})$ are obtained as best linear unbiased predictors. The fitted linear trajectory generated by LRF--adaptiveLMM for subject $i$ at node $\mathcal{I}$ is therefore
\begin{equation}
\widehat{\mathbf{Y}}_{i,\mathcal{I}}(t)
=
\widehat{\beta}_{0,\mathcal{I}}
+ \widehat{\beta}_{t,\mathcal{I}}\, t
+ \mathbf{X}_{i,\mathcal{I}}^{\top} \widehat{\boldsymbol{\beta}}_{\mathcal{I}}
+ \widehat{b}_{i0,\mathcal{I}}
+ \widehat{b}_{i1,\mathcal{I}}\, t .
\label{eq:lmm_fitted_curve}
\end{equation}

In summary, for the semiparametric LRF--adaptiveLMM, the covariate set $\mathbf{X}_{i,\mathcal{I}}$ is explicitly incorporated into the node-specific longitudinal trajectory estimation $\widehat{\mathbf{Y}}_{i,\mathcal{I}}(t)$ through the fixed-effects component in model \eqref{eq:lmm_fitted_curve}. However, LRF--PACE remains fully nonparametric at the node level: covariates are not explicitly incorporated into the construction of $\widehat{\mathbf{Y}}_{i,\mathcal{I}}(t)$ in model \eqref{eq:pace_fitted_curve}; instead, they influence the longitudinal trajectory estimation implicitly through the covariate-driven tree partitioning process.

\subsection{Splitting Rule}\label{sec:splitting_rule}

After subject-level smoothed trajectories 
$\{\widehat{\mathbf{Y}}_{i,\mathcal{I}}(t),~~i\in\mathcal{I}\}$
are estimated using a node-wise longitudinal smoother
(LRF--PACE in \eqref{eq:pace_fitted_curve} or
LRF--adaptiveLMM in \eqref{eq:lmm_fitted_curve}),
the representative trajectory at node $\mathcal{I}$ is defined as
\begin{equation}
\widehat{\mathbf{f}}_{\mathcal{I}}(t)
=
\frac{1}{n_{\mathcal{I}}}
\sum_{i \in \mathcal{I}} \widehat{\mathbf{Y}}_{i,\mathcal{I}}(t),
\label{eq:node_rep_traj}
\end{equation}
where $n_{\mathcal{I}}$ denotes the number of subjects at node $\mathcal{I}$. 

Consider a candidate split at node $\mathcal{I}$ on covariate $X_k$, where $k \in \{1,\dots,p\}$. Let $X_{ik}$ denote the value of $X_k$ for subject $i$. Let $c$ denote a candidate threshold. For continuous covariates, $c$ is a candidate quantile within the domain of $X_k$, while for categorical covariates, $c$ represents each level of the category of $X_k$. The node ${\mathcal{I}}$ is divided into two child nodes
\[
\mathcal{I}_L = \{\, i \in \mathcal{I} : X_{ik} \le c \,\},
\qquad
\mathcal{I}_R = \{\, i \in \mathcal{I} : X_{ik} > c \,\},
\]
with sample sizes $n_{\mathcal{I}_L}$ and $n_{\mathcal{I}_R}$, respectively.

Let $\widehat{\mathbf{f}}_{\mathcal{I}_L}(t)$ and $\widehat{\mathbf{f}}_{\mathcal{I}_R}(t)$ denote the representative trajectories of the left and right child nodes obtained via Equation~\eqref{eq:node_rep_traj}. To separate heterogeneous trajectories, we propose the trajectory-based splitting criterion
\begin{equation}
\mathcal{S}(X_k, c)
=
\frac{n_{\mathcal{I}_L}\, n_{\mathcal{I}_R}}
     {n_{\mathcal{I}_L} + n_{\mathcal{I}_R}}
\int_{\mathcal{T}}
\bigl\{
\widehat{\mathbf{f}}_{\mathcal{I}_L}(t)
-
\widehat{\mathbf{f}}_{\mathcal{I}_R}(t)
\bigr\}^2 dt ,
\label{eq:trajectory_split_criterion}
\end{equation}
where $\mathcal{T}$ denotes the corresponding time domain.

This criterion promotes splits that maximize separation between two child nodes by measuring the integrated squared difference between their representative trajectories, capturing both magnitude and trajectory discrepancy. The weighting factor $n_{\mathcal{I}_L} n_{\mathcal{I}_R}/(n_{\mathcal{I}_L} + n_{\mathcal{I}_R})$ penalizes those highly unbalanced splits, preventing partitions driven by a small number of noisy subjects.

The optimal split at node $\mathcal{I}$ is selected by maximizing $\mathcal{S}(X_k, c)$ across all candidate covariates and thresholds:
\[
(X_{k^*}, c^*)
=
\arg\max_{k,\,c}\; \mathcal{S}(X_k, c).
\]
Recursive partitioning continues while a node remains eligible for further splitting. When the sample size at a node is too small, the estimated trajectory becomes overly sensitive to individual observations, increasing the risk of splits driven by noise rather than meaningful differences in longitudinal response trajectories. To control this, we impose two tuning parameters: the minimum node size $n_{\min}$, such that each resulting child node must contain sufficient numbers of subjects. However, the minimum node size constraint alone may still allow excessively deep trees. Therefore, tree growth is restricted by a maximum depth constraint, $d_{\max}$, to prevent overfitting.

\subsection{Prediction for New Subjects and Forecast for Future Time Points}
\label{sec:prediction}

We partition the data into a training set $\mathcal{D}_{\mathrm{train}}$ and a test set $\mathcal{D}_{\mathrm{test}}$, with sample sizes $N_{\mathrm{train}}$ and $N_{\mathrm{test}}$, respectively. The LRF model is fitted using $\mathcal{D}_{\mathrm{train}}$, and the predictive performance is evaluated on $\mathcal{D}_{\mathrm{test}}$.

The proposed LRF enables two prediction settings. First, given only the covariates for a new subject $i_{\mathrm{new}} \in \mathcal{D}_{\mathrm{test}}$, we predict its entire trajectory over the study period. Second, for an existing subject $i \in \mathcal{D}_{\mathrm{train}}$ with observed longitudinal measurements, we forecast unknown responses at future time beyond the subject’s study period.

After a tree is constructed, a new subject $i_{\mathrm{new}} \in \mathcal{D}_{\mathrm{test}}$ is routed through the $b$th tree according to the splitting rules based on its covariates $\mathbf{X}_{i_\mathrm{new}}$ and assigned to a terminal node $\mathcal{I}_{i_\mathrm{new}}$. 
The tree-level predicted trajectory for the new subject is then obtained as
\begin{equation}
\widetilde{\mathbf{Y}}_{i_\mathrm{new}}^{(b)}(t)
=
\begin{cases}
\widehat{\mathbf{f}}_{\mathcal{I}_{i_\mathrm{new}}}(t), & \text{LRF--PACE}, \\[4pt]
\widehat{\mathbf{Y}}_{i_\mathrm{new},\mathcal{I}_{i_\mathrm{new}}}^{\,\mathrm{FE}}(t), 
& \text{LRF--adaptiveLMM}.
\end{cases}
\label{eq:lrf_tree_pred_new}
\end{equation}
For LRF--PACE, $\widehat{\mathbf{f}}_{\mathcal{I}_{i_\mathrm{new}}}(t)$ is defined by Equation~\eqref{eq:node_rep_traj}.
For LRF--adaptiveLMM, the new-subject prediction is based only on the fixed-effects component because subject-specific random effects cannot be estimated without observed longitudinal measurements. Specifically, $\widehat Y_{i_{\mathrm{new}},\mathcal{I}_{i_\mathrm{new}}}^{\,\mathrm{FE}}(t)$ denotes the fixed-effects prediction given by
\[
\widehat{\mathbf{Y}}_{i_{\mathrm{new}},\mathcal{I}_{i_\mathrm{new}}}^{\,\mathrm{FE}}(t)
=
\widehat{\beta}_{0,\mathcal{I}_{i_\mathrm{new}}}
+ \widehat{\beta}_{t,\mathcal{I}_{i_\mathrm{new}}}\, t
+ \mathbf{X}_{i_{\mathrm{new}},\mathcal{I}_{i_\mathrm{new}}}^\top 
\widehat{\boldsymbol\beta}_{\mathcal{I}_{i_\mathrm{new}}}.
\]

For future-time forecasting, consider an existing subject $i \in \mathcal{D}_{\mathrm{train}}$ with observed longitudinal responses at time points $t_{i1}<\cdots<t_{iT_i}$ and a future time point $t_{\mathrm{new}}>t_{iT_i}$. The subject is routed through the $b$th tree based on its covariates and assigned to a terminal node $\mathcal{I}$. Its tree-level forecast for future time is obtained as
\begin{equation}
\widetilde{\mathbf{Y}}_i^{(b)}(t_{\mathrm{new}})
=
\begin{cases}
\widehat{\mathbf{f}}_{\mathcal{I}}(t_{\mathrm{new}}), & \text{LRF--PACE}, \\
\widehat{\mathbf{Y}}_{i,\mathcal{I}}(t_{\mathrm{new}}), & \text{LRF--adaptiveLMM}.
\end{cases}
\label{eq:lrf_tree_pred_future}
\end{equation}
For LRF--PACE, $\widehat{\mathbf{f}}_{\mathcal{I}}(t)$ is again derived from Equation~\eqref{eq:node_rep_traj}. For LRF--adaptiveLMM, $\widehat{\mathbf{Y}}_{i,\mathcal{I}}(t)$ is obtained by incorporating both fixed effects and the estimated subject-specific random effects, as in Equation~\eqref{eq:lmm_fitted_curve}. The corresponding forecast is then obtained by evaluating Equation~\eqref{eq:lrf_tree_pred_future} at a future time point $t_{\mathrm{new}}$.

After aggregating all $B$ trees, the final prediction generated by LRF for new subjects is given by
\begin{equation}
\label{eq:lrf_aggregation_newsbj}
\begin{aligned}
\widetilde{\mathbf{Y}}_{i_{\mathrm{new}}}(t)
&=
\frac{1}{B}\sum_{b=1}^B
\widetilde{\mathbf{Y}}_{i_{\mathrm{new}}}^{(b)}(t),
\end{aligned}
\end{equation}
and the forecast for future time generated by LRF is given by
\begin{equation}
\label{eq:lrf_aggregation_newt}
\begin{aligned}
\widetilde{\mathbf{Y}}_i(t_{\mathrm{new}})
&=
\frac{1}{B}\sum_{b=1}^B
\widetilde{\mathbf{Y}}_i^{(b)}(t_{\mathrm{new}}).
\end{aligned}
\end{equation}

\subsection{Permutation Variable Importance Measure}\label{sec:PVIM}
The LRF framework assesses the importance of each covariate through a permutation variable importance measure computed using out-of-bag (OOB) samples. Let $\mathcal{O}_b \subset \mathcal{D}_{\mathrm{train}}$ denote the OOB sample associated with the $b$th tree. For each subject $i \in \mathcal{O}_b$, its predicted discrete response values $\widetilde{\mathbf{Y}}_{i}^{(b)}(t_{ij})$ at each tree are obtained by interpolating the observed time sampling $\{t_{ij}\}_{j=1}^{T_i}$ through Equation~\eqref{eq:lrf_tree_pred_new}.

For a given covariate $X_k$, its values are randomly permuted across subjects within $\mathcal{O}_b$, while all other covariates remain unchanged. Let $\pi(k)$ denote a permutation applied to covariate $X_k$. The permutation variable importance measure for $X_k$ is defined as
\begin{equation}
\begin{split}
\mathrm{PVIM}(X_k)
&=
\frac{1}{B}
\sum_{b=1}^B
\left[
\frac{1}{|\mathcal{O}_b|}
\sum_{i\in\mathcal{O}_b}
\frac{1}{T_i}
\sum_{j=1}^{T_i}
\right. \\
&\qquad\left.
\left\{
\bigl(Y_{ij}-\widetilde{\mathbf{Y}}_{i,\pi(k)}^{(b)}(t_{ij})\bigr)^2
-
\bigl(Y_{ij}-\widetilde{\mathbf{Y}}_{i}^{(b)}(t_{ij})\bigr)^2
\right\}
\right].
\end{split}
\label{eq:pvim_forest}
\end{equation}
where $\widetilde{\mathbf{Y}}_{i}^{(b)}(t)$ and $\widetilde{\mathbf{Y}}_{i,\pi(k)}^{(b)}(t)$ denote the predictions obtained from the $b$th tree before and after permuting $X_k$, respectively. Thus, the PVIM is the average change in OOB prediction error across all trees after permuting $X_k$. Larger values of $\mathrm{PVIM}(X_k)$ indicate that permuting $X_k$ substantially increases prediction error, implying a greater contribution of $X_k$ to longitudinal trajectory prediction. The PVIM is therefore utilized to rank covariates within the LRF framework for variable selection purpose.

\subsection{Frequency-based Interaction Importance}
\label{sec:interaction_exp}

A high PVIM value does not distinguish whether a covariate's importance arises from its individual effect or from interactions with other covariates. To address this limitation, we propose a frequency-based strategy that leverages the tree structures of the LRF to quantify interaction importance. The key idea is to examine the co-occurrence frequencies of multi-way covariate combinations along root-to-terminal paths across all trees and all cross-validation replications. Specifically, for each tree, we trace every root-to-terminal path and record all subsets of covariates that are involved for splitting along every path. For example, for a three-node tree path with splits based on covariates $\{x_2, x_1, x_5\}$, we record one count for each nested interaction subset $\{x_2\}$, $\{x_1, x_2\}$, and $\{x_1, x_2, x_5\}$, where the order of covariates within each subset does not matter. Repeating this process across all paths yields co-occurrence counts for all occurred interaction subsets. Covariates that appear near the root partition a larger portion of the subjects and therefore influence more subjects than those appearing near the terminal node. As a result, this rule assigns greater frequency to the covariates near the root nodes than those near the terminal nodes. To improve stability, the LRF model is fitted across ten cross-validation replications, and interaction counts are aggregated over all trees from all replications. The resulting aggregated co-occurrence frequencies produce an overall ranked list spanning two-way, three-way, four-way, and higher-way interactions, allowing interactions of all finite orders to be identified within one procedure.

\subsection{Tuning Parameters}

Several tuning parameters need to be specified to implement the LRF framework. The first is \texttt{mtry}, which determines the number of candidate covariates randomly considered at each split. Following common practice in random forests, we set \texttt{mtry} to be $p/3$. The second parameter is the minimum node size, $n_{\min}$, which is set to $n_{\min}=20$ to prevent terminal nodes from becoming too small. The third parameter is the maximum tree depth, $d_{\max}$, which balances model complexity and predictive performance by limiting the number of nodes along any root-to-terminal path. It is preselected by minimizing the fivefold cross-validation error. The final tuning parameter is the number of trees, $B$, in the forest. We use $B=50$, which is sufficient to yield stable predictive performance, as demonstrated in the simulation studies in Section~\ref{sec:simulation}. For the LRF--PACE implementation, one additional tuning parameter is required: the number of FPCs, $L$, which is selected to explain at least 90\% of the total variation at each node.

The LRF algorithm is implemented in \textsf{R} with support for parallel computing to facilitate efficient training of large ensembles. A complete workflow of the LRF procedure is provided in Algorithm~\ref{alg:LRF}.

\begin{algorithm}
\caption{\enskip Longitudinal Random Forest}\label{alg:LRF}
\begin{algorithmic}

\State \textbf{Input:} Covariates $\{\mathbf{X}_i\}_{i=1}^{N}$,
longitudinal observations $\{(t_{ij},Y_{ij})\}$, node-wise smoother (PACE or adaptiveLMM), and tuning parameters: $B$,
\texttt{mtry}, $n_{\min}$, $d_{\max}$\\

\For{$b=1,\ldots,B$}
  \State draw a bootstrap sample and OOB sample\\
 
  \While{a node $\mathcal{I}$ satisfies the splitting conditions $n_{\min}$ and $d_{\max}$}
    \State randomly select \texttt{mtry} candidate covariates

    \For{each candidate split $(X_k,c)$}

      \State estimate $\widehat{\mathbf{Y}}_{i,\mathcal{I}}(t)$ using Equation \eqref{eq:pace_fitted_curve} or \eqref{eq:lmm_fitted_curve}
      \State compute $\widehat{\mathbf{f}}_{\mathcal{I}(t)}$ using
      Equation \eqref{eq:node_rep_traj}
      \State compute $\mathcal{S}(X_k,c)$ using
      Equation \eqref{eq:trajectory_split_criterion}
    \EndFor \\

    \State select $(X_{k^*},c^*)$ that maximizes $\mathcal{S}(X_k,c)$
    \State partition $\mathcal{I}$ into $\mathcal{I}_L$ and
    $\mathcal{I}_R$
  \EndWhile \\

\EndFor \\

\State \textbf{Prediction and Forecast:}

\For{each tree $b$} 
\State route a subject to a terminal node 
\State obtain $\widetilde{\mathbf{Y}}_{i_\mathrm{new}}^{(b)}(t)$ using Equation~\eqref{eq:lrf_tree_pred_new} or $\widetilde{\mathbf{Y}}_i^{(b)}(t_{\mathrm{new}})$ using Equation~\eqref{eq:lrf_tree_pred_future} 
\EndFor 
\State aggregate tree-level predictions using Equation~\eqref{eq:lrf_aggregation_newsbj} or \eqref{eq:lrf_aggregation_newt}\\

\State \textbf{Variable Importance:}

\For{each covariate $X_k$}
  \State compute $\mathrm{PVIM}(X_k)$ using
  Equation~\eqref{eq:pvim_forest}
\EndFor\\

\State \textbf{Interaction Exploration:}
\State aggregate covariate co-occurrence frequencies across all
root-to-terminal paths\\

\State \textbf{Output:} Fitted LRF ensemble, trajectory predictions/forecast, PVIM scores, and multi-way covariate co-occurrence frequencies

\end{algorithmic}
\end{algorithm}

\section{Simulation Studies}\label{sec:simulation}

We evaluate the finite-sample performance of the proposed Longitudinal Random Forest methods under two data-generating settings. Setting (i) considers a fully nonparametric mechanism in which response trajectories are simulated as functions and complex covariate effects are introduced through a tree-structured model. Setting (ii) generates data to closely mimic the real data, INSULIN trial, to reproduce its exact missingness pattern. 

\subsection{Evaluation Metrics}

We evaluate variable selection and prediction/forecast performance over 100 simulation replicates  using the following two criteria:

\begin{itemize}

\item \textbf{Average Selection Rank}: 
The average rank of each true covariate based on the permutation variable importance measure. Smaller ranks indicate more accurate variable selection performance.

\item \textbf{Prediction Error (PE)}:
We evaluate new-subject prediction accuracy by the mean squared prediction error:
\begin{equation}
\label{eq:pe_both}
\mathrm{MSPE}_{\mathrm{new \ subject}}
=
\frac{1}{N_{\mathrm{test}}}
\sum_{i_{new} \in \mathcal{D}_{\mathrm{test}}}
\frac{1}{T_{i_{new}}}
\sum_{j=1}^{T_{i_{new}}}
\left\{
Y_{i_{new}j}
-
\widetilde{\mathbf{Y}}_{i_{new}}(t_{i_{new}j})
\right\}^2. \\
\end{equation}
$\mathrm{MSPE}_{\mathrm{new \ subject}}$ measures the discrepancy between predicted and observed response values across all observed time points. Smaller MSPE values indicate higher prediction accuracy. 
\end{itemize}

\subsection{Simulation 1}

We first generate functional data with regular and dense observations for $N = 300$ subjects observed at $T = 20$ equally spaced time points between 0 and 10. A smooth function is created through a functional principal component analysis (FPCA) structure with a population mean function and two eigenfunctions as follows,
\[
\begin{aligned}
\mu(t) &= t/10 + \sin(t),\\
\phi_1(t) &= -\frac{\cos(\pi t/10)}{\sqrt{5}},\\
\phi_2(t) &= \frac{\sin(\pi t/10)}{\sqrt{5}}.
\end{aligned}
\]
Baseline covariates are generated as $\mathbf{X}_i \sim \mathcal{N}_p(0,\Sigma_\mathbf{X})$ with $p=100$, where $\Sigma_{\mathbf X}$ follows an AR(1) covariance structure with $(\Sigma_{\mathbf X})_{k_1k_2} = \rho^{|k_1-k_2|}$ with $\rho = 0.2$. 

To induce heterogeneous trajectories associated with nonlinear main effects and interactions, subject-specific FPC scores are generated through a randomized tree-based mechanism. Let $S_{\mathrm{true}}=\{X_{1},X_{11},X_{21}\}$ denote the set of true covariates. We first generate two node-specific FPC score vectors,
$\boldsymbol{\xi}^{L} = (\xi_{1}^{L},\xi_{2}^{L})^\top \sim N\!\left( \mathbf 0, \mathrm{diag}(4,1) \right)$ and $\boldsymbol{\xi}^{R} = (\xi_{1}^{R},\xi_{2}^{R})^\top \sim N\!\left( \mathbf 0, \mathrm{diag}(3,2) \right)$, which are used for all trees throughout the simulation. Next, $B=10$ trees are generated independently. For each tree, one to three covariates are randomly drawn without replacement from $S_{\mathrm{true}}$. Each split is determined by a simple threshold rule of the form $X_k < 0.5$. At each node, a subject is assigned to the corresponding node-specific FPC score vector, $\boldsymbol{\xi}^{L}$ or $\boldsymbol{\xi}^{R}$, depending on whether the subject is assigned to the left or right child node. The tree-specific FPC score vector is defined as the average of all node-specific FPC score vectors along the subject's root-to-terminal path. For example, consider a tree consisting of only two decision nodes, corresponding to variables $X_1$ and $X_{11}$. Its tree-specific FPC score vector for subject $i$ is given by
\[
\begin{aligned}
\boldsymbol{\xi}_i^{(b)}
&=
\frac{1}{2}
\Bigl\{
\mathbf{1}\{X_{i1}<0.5\}\boldsymbol{\xi}^{L}
+
\mathbf{1}\{X_{i1}\ge 0.5\}\boldsymbol{\xi}^{R}
\\
&\qquad\quad
+
\mathbf{1}\{X_{i11}<0.5\}\boldsymbol{\xi}^{L}
+
\mathbf{1}\{X_{i11}\ge 0.5\}\boldsymbol{\xi}^{R}
\Bigr\},
\end{aligned}
\]
where $\mathbf{1}\{\cdot\}$ denotes the indicator function, which equals 1 when the specified condition is satisfied and 0 otherwise.

Finally, the subject-specific FPC score vector is obtained by averaging the tree-specific FPC score vectors across all $B=10$ trees plus independent Gaussian perturbations,
\[
\widetilde{\boldsymbol{\xi}}_i
=
\frac{1}{B}
\sum_{b=1}^{B}
\boldsymbol{\xi}_i^{(b)}
+
\boldsymbol{\varepsilon}_i,
\qquad
\boldsymbol{\varepsilon}_i
\sim
N(\mathbf{0},\operatorname{diag}(1,1)).
\]
The simulated dense functional trajectory for subject $i$ is generated as
\begin{equation}
\label{eq:s2_fun}
g_i(t)
=
\mu(t)
+
\widetilde{\xi}_{i1}\phi_1(t)
+
\widetilde{\xi}_{i2}\phi_2(t)
+
f(X_{i1},X_{i11},X_{i21}),
\end{equation}
where $f(\cdot)$ introduces additional nonlinear and interactive covariate effects if needed. For example,
\[
f(X_{1},X_{11},X_{21})
=
X_{1}
+
X_{11}^2
+
X_{1}X_{21}.
\]

After generating densely and regularly sampled functional trajectories at $T$ equally spaced time points, we set sparse and irregular time sampling $t_{ij}$ by randomly deleting observations from each subject’s trajectory with missing rate $r \in \{30\%, 50\%, 70\%\}$, representing low, moderate, and severe levels of missingness. Note that under a missing rate as high as $r = 70\%$, each subject may retain only 4–8 observed time points, resulting in extremely sparse longitudinal response measurements. The discrete observations are generated as
\[
Y_{ij} = g_i(t_{ij}) + \varepsilon_{ij}, 
\qquad \varepsilon_{ij} \sim {N}(0,1),
\]
where $g_i(t)$ is generated in equation \eqref{eq:s2_fun}. Subjects are then randomly split into $N_{\mathrm{train}} = 200$ for model fitting and $N_{\mathrm{test}} = 100$ for prediction.

\begin{table}[!t]
\centering
\caption{Average importance ranks of the true covariates in Simulation 1 under three missingness rates. Smaller ranks indicate more accurate variable selection performance.}
\label{tab:s2-fs}
\vspace{0.5cm}
\small
\begin{tabular}{lccccc}
\toprule
{Missing Rate} & {Method} & \multicolumn{3}{c}{Average Selection Rank} \\
\cmidrule(lr){3-5}
 &  & $X_1$ & $X_{11}$ & $X_{21}$ \\
\midrule
30\% & SPLF  & \textbf{1.00} & \textbf{2.00} & 3.81 \\
     & SREEMF     & \textbf{1.00} & 2.02 & 3.95 \\
     & LRF--adaptiveLMM  & 1.16 & 2.33 & 7.43 \\
     & LRF--PACE & \textbf{1.00} & \textbf{2.00} & \textbf{3.55} \\
\midrule
50\% & SPLF   & \textbf{1.00} & 2.27 & 4.74 \\
     & SREEMF     & \textbf{1.00} & 2.01 & 4.01 \\
     & LRF--adaptiveLMM  & 1.36 & 2.59 & 8.35 \\
     & LRF--PACE & \textbf{1.00} & \textbf{2.00} & \textbf{3.67} \\
\midrule
70\% & SPLF   & 1.24 & 8.20 & 25.51 \\
     & SREEMF     & \textbf{1.00} & {2.06} & 4.38 \\
     & LRF--adaptiveLMM  & 1.45 & 2.88 & 9.66 \\
     & LRF--PACE & \textbf{1.00} & \textbf{2.02} & \textbf{4.11} \\
\bottomrule
\end{tabular}
\end{table}

\begin{table}[ht]
\centering
\caption{Average prediction errors for Simulation~1 across 100 replications: predicting the entire response trajectories of new subjects under three missingness scenarios derived from $T=20$. Smaller values indicate higher predictive accuracy.}
\label{tab:s2-pde}
\vspace{0.5cm}

\resizebox{\textwidth}{!}{%
\begin{tabular}{lccccccc}
\toprule
Missing Rate & SPLF & SREEMF & LongCART & LMM & GEE & LRF--adaptiveLMM & LRF--PACE \\
\midrule
30\% & \textbf{1.66} & 1.89 & 2.26 & 2.05 & 1.91 & 1.79 & 1.73\\
50\% & 1.77 & 1.99 & 2.27 & 2.68 & 2.65 & 1.83 & \textbf{1.76}\\
70\% & 1.97 & 2.02 & 2.28 & 2.79 & 2.78 & 1.97 & \textbf{1.81}\\
\bottomrule
\end{tabular}%
}
\end{table}

Table \ref{tab:s2-fs} summarizes the average selection ranks for the three true covariates over 100 replications under missingness rates of 30\%, 50\%, and 70\%, respectively. For the easiest case, $X_1$, all methods perform nearly perfect under all three missingness rates, with average selection ranks close to 1. For the moderate case, $X_{11}$, most methods still maintain accurate and stable selection, with average ranks around 2 to 3. The two LRF variants and SREEMF remain robust even under 70\% missingness, whereas SPLF’s average selection rank increases to 8.20 under severe sparsity. The largest performance differences arise for the hardest case, $X_{21}$. LRF--PACE consistently yields the smallest average selection ranks across all missingness rates, with average ranks ranging from 3 to 4, which aligns well with the number of true covariates. SREEMF also performs competitively for $X_{21}$, although its average ranks are slightly higher than those of LRF--PACE. At the highest missingness level of 70\%, SPLF produces the worst average selection rank of 25.51 for $X_{21}$. 

Table \ref{tab:s2-pde} reports the average prediction error for new test subjects over 100 replications under missingness rates of 30\%, 50\%, and 70\%, respectively. LRF--PACE achieves the smallest prediction errors at moderate and high missingness levels (PE = 1.76 and 1.81). LRF–adaptiveLMM also performs well and shows stable prediction performance as missingness increases, with prediction error rising only modestly from 1.79 to 1.97. SPLF achieves a slightly lower prediction error (PE = 1.66) than LRF--PACE at the lowest missingness level of 30\%, but its prediction error increases as sparsity becomes severe. Classical LMM, GEE, and LongCART yield the largest prediction errors in all settings.

\subsection{Simulation 2}\label{subsec:lrf-real}

This simulation setting is designed to closely mimic the time-specific missingness pattern observed in the INSULIN trial, thereby providing a realistic assessment of how different longitudinal analysis methods perform in this unusually sparse setting. We generate $N=300$ subjects observed at $M=24$ discrete time points, each with $p=20$ covariates $\mathbf{X}_i \sim \mathcal{N}_p(\mathbf{0},\Sigma_X)$, where $(\Sigma_X)_{k_1k_2}=\rho^{|k_1-k_2|}$ with $\rho=0.2$ and unit variances, for $k_1,k_2=1,\ldots,20$. Let $S_{\mathrm{true}}=\{X_{1},X_{2},X_{3}\}$ denote the set of true covariates. For subject $i$, the underlying trajectory is generated as
\begin{equation}
\label{eq:s3_fun}
\begin{aligned}
g_i(t)
&=
u_i\bigl(1-e^{-v_i t}\bigr)
-
0.4\,w_i
\sin\!\left(\frac{1.2\pi t}{24}\right)
\\
&\quad
+
0.5X_{i1}
-
0.2X_{i2}
-
X_{i1}X_{i3}
+
X_{i2}X_{i3}.
\end{aligned}
\end{equation}
with
\[
u_i \sim \mathrm{Unif}(-4,0),
\qquad
v_i \sim \mathrm{Unif}(0,5),
\qquad
w_i \sim \mathrm{Unif}(0,2).
\]
The first term, $u_i(1-e^{-v_i t})$, generates a smooth nonlinear decreasing trend whose magnitude and rate of change vary across subjects through $u_i$ and $v_i$. The second term introduces additional oscillatory behavior that allows trajectories to exhibit different degrees of curvature and temporal fluctuation. The remaining terms incorporate nonlinear covariate effects and interactions, allowing subjects with different baseline characteristics to follow systematically different trajectory patterns and thereby introducing additional between-subject heterogeneity.

To closely resemble the irregular and sparse sampling schedules observed in the real dataset, we impose the empirical week-specific missingness rates estimated directly from the INSULIN trial data (see Section~\ref{sec:data}). Specifically, missingness is $0\%$ at Week~1, alternates between $97\%$ for Weeks~2 and~4 and $3\%$ for Weeks~3 and~5, ranges from $45\%$ to $75\%$ between Weeks 6 and 14, reaches $100\%$ for all weeks between Weeks 15 and 22 except Week 20, which has a missingness rate of $98.8\%$, and is approximately $65\%$ at Weeks 23 and 24, toward the end of the study period. To generate the discrete and irregular time sampling $t_{ij}$, time points at each week are randomly removed according to the corresponding empirical missingness rate. The discrete sparse and irregular longitudinal responses are then generated by evaluating $g_i(t)$ in equation \eqref{eq:s3_fun} at $t_{ij}$ and adding measurement noise,
\[
Y_{ij}
=
g_i(t_{ij})
+
\varepsilon_{ij},
\qquad
\varepsilon_{ij}
\sim
N(0,1).
\]
Subjects are then randomly split into a training set of size $N_{\mathrm{train}}=200$ and a test set of size $N_{\mathrm{test}}=100$.

In particular, this design reproduces the large empty region between Weeks 15 and 22 present in the INSULIN trial. Such an extended interval with almost no observations poses a substantial challenge for nonparametric trajectory estimation because smoothing methods rely on sufficient local information. In exploratory analyses, the original LRF--PACE model did not perform well under the entire 24-week study period. To accommodate this nearly empty region, we introduce a modified version, denoted LRF--PACE--2, which employs an adaptive piecewise smoothing strategy. Specifically, LRF--PACE is applied only to Weeks 1 to 14, where sufficient observations are available for reliable estimation. The estimated trajectory is then extrapolated through Weeks 15 to 22. For the final period, Weeks 23--24, it contains only two scheduled visits, therefore, trajectory values are estimated by simply averaging the available observations. This modification is only specific to LRF--PACE because its trajectory estimation relies on nonparametric smoothing. In contrast, LRF--adaptiveLMM is based on a parametric linear mixed-effects model and can naturally generate predictions at time points with no observations through model-based interpolation.

\begin{table}[!t]
\centering
\caption{Average importance ranks of the true covariates in Simulation 2. Smaller ranks indicate more accurate variable selection performance.}
\label{tab:s3-fs}
\vspace{0.5cm}
\small
\begin{tabular}{lccc}
\toprule
Method & $X_1$ & $X_{2}$ & $X_{3}$ \\
\midrule
SPLF             & 1.75 & 4.46 & 9.75 \\
SREEMF           & 1.09 & 2.29 & 4.23 \\
LRF--adaptiveLMM         & 1.07 & 2.52 & 5.85 \\
LRF--PACE        & 1.90 & 4.17 & 6.68 \\
\textbf{LRF--PACE--2} & \textbf{1.01} & \textbf{2.15} & \textbf{3.94} \\
\bottomrule
\end{tabular}
\end{table}

\begin{table}[!t]
\centering
\caption{Average prediction errors for Simulation~2 across 100 replications: predicting the entire response trajectories of new subjects over Weeks 1-24 under the empirical missingness pattern derived from the INSULIN data. Smaller values indicate higher predictive accuracy.}
\label{tab:s3-pde}
\vspace{0.5cm}
\resizebox{\columnwidth}{!}{%
\begin{tabular}{cccccccc}
\toprule
SPLF & SREEMF & LongCART & GEE & LMM & LRF--adaptiveLMM & LRF--PACE & LRF--PACE--2\\
\midrule
3.24 & 3.57 & 3.93 & 4.05 & 4.20 & 3.47 & 3.28 & \textbf{3.06}\\
\bottomrule
\end{tabular}
}
\end{table}

Table \ref{tab:s3-fs} summarizes the average selection ranks of the three true covariates across 100 replications in Simulation 2. For the easiest case, $X_1$, all methods achieve nearly perfect variable selection performance, with average selection ranks between 1 and 2. Differences become noticeable for the moderate case, $X_2$, where LRF--PACE--2 achieves the smallest average selection rank (2.15), followed closely by SREEMF (2.29) and LRF--adaptiveLMM (2.52). The largest performance difference appears at the hardest case $X_3$. LRF--PACE--2 is the only method attaining an average selection rank below 4 (3.94), outperforming all competing approaches. SREEMF provides the second-best performance (4.23). LRF--PACE yields a high selection rank (6.68), while SPLF performs the worst overall, with the largest average selection rank (9.75). Overall, these results highlight the necessity of the adaptive piecewise smoothing strategy in LRF--PACE--2 for handling the large empty gap present in the data.

Table~\ref{tab:s3-pde} reports the average prediction error for the estimated response trajectories over Weeks~1--24 for new subjects in the test data across 100 replications in Simulation~2. Among all methods, LRF--PACE--2 attains the lowest prediction error (PE = 3.06), indicating the highest  accuracy for new subject prediction. LRF--adaptiveLMM (PE = 3.47) and SPLF (PE = 3.24) also perform competitively. LongCART, the classical LMM, and GEE yield the largest prediction errors.

We next assess each model's ability to forecast unobserved responses at future time points for existing training subjects. In practice, the response values at future time points are not observed, motivating the need for forecasting. However, forecasting performance cannot be evaluated directly without knowing the actual future observations. Therefore, for evaluation purposes, each model is fitted over Weeks 1--10 and then predicts responses at Weeks 11--14, which are pretended as future time points. 

Let $\mathcal{T}_{\mathrm{new}}$ denote the future-time domain, and let $t_{ij}^{\mathrm{new}}\in\mathcal{T}_{\mathrm{new}}$ denote the $j$th future evaluation time point for subject $i$, corresponding to the future time point $t_{\mathrm{new}}$ defined in Section~\ref{sec:prediction}. Let $Y_{ij}^{\mathrm{new}}$ denote the response value at $t_{ij}^{\mathrm{new}}$, and let $T_{i,\mathrm{new}}$ be the number of future evaluation time points for subject $i$. Future-time forecasting accuracy is quantified by the mean squared prediction error
\begin{equation}
\label{eq:mspe_future}
\mathrm{MSPE}_{\mathrm{future}}
=
\frac{1}{N_{\mathrm{train}}}
\sum_{i\in\mathcal{D}_{\mathrm{train}}}
\frac{1}{T_{i,\mathrm{new}}}
\sum_{j=1}^{T_{i,\mathrm{new}}}
\left\{
Y_{ij}^{\mathrm{new}}
-
\widetilde{\mathbf{Y}}_{i}
\bigl(t_{ij}^{\mathrm{new}}\bigr)
\right\}^{2}.
\end{equation}
The $\mathrm{MSPE}_{\mathrm{future}}$ measures the discrepancy between the forecasted and observed responses at future evaluation time points. Smaller MSPE values indicate higher forecasting accuracy. The resulting forecasting performance is reported in Table~\ref{tab:s4-pde}.

\begin{table}[htbp]
\centering
\caption{Average forecast errors for Simulation~2 across 100 replications: forecasting future response trajectories of existing subjects over Weeks~11--14 by training observations over Weeks~1--10 under the empirical missingness pattern derived from the INSULIN trial data. Lower values indicate higher forecast accuracy.}
\label{tab:s4-pde}
\vspace{0.5cm}
\small
\begin{tabular}{lccccc}
\toprule
SPLF & SREEMF & LongCART & LRF--adaptiveLMM & LRF--PACE\\
\midrule
 3.25 & 4.24 & 4.19 &  2.95 & \textbf{2.09}\\
\bottomrule
\end{tabular}
\end{table}

Among all methods, LRF--PACE achieves the lowest forecasting error (2.09), demonstrating its strong ability to forecast dynamic temporal trends beyond the observed time window. LRF--adaptiveLMM (2.95), also exhibits strong forecasting performance and outperforms the other competing approaches, including SPLF (3.25), SREEMF (4.24), and LongCART (4.19). These results highlight the benefits of the LRF framework, in which the tree structure recursively partitions subjects into subgroups whose members share homogeneous temporal trends. This process minimizes within-node trajectory variation, thereby stabilizing node-level estimation and improving both prediction and forecasting accuracy. 

\section{Real Data Analysis}\label{sec:realdata}
We apply the proposed LRF models to the motivating INSULIN trial real dataset \citep{buse2009durable, buse2011durable}. The response consists of longitudinal HbA1c measurements collected up to 24 weeks. The baseline covariates are the 19 clinical and biochemical variables described in Section~\ref{sec:data}. Among these covariates, baseline HbA1c is included in all analyses, following current regulatory guidance from both the U.S. Food and Drug Administration (FDA) and the European Medicines Agency (EMA), which regard baseline adjustment as appropriate \citep{FDA2023Covariates,EMA2015Covariates}. The 1,520 eligible subjects are randomly divided into a training set of 1,000 subjects and a test set of 520 subjects. This random splitting is repeated 10 times as a standard cross-validation procedure, and model performance is averaged across the resulting replications.

In the real data setting, comparisons across methods focus primarily on prediction performance because the truly influential covariate effects are unknown. Therefore, variable importance results are reported primarily for interpretation rather than for comparison. Although SREEMF performs well in the simulation studies, its publicly available R implementation could not be successfully applied to the INSULIN dataset because of numerical errors, likely arising from singularity issues during matrix estimation. Consequently, SREEMF is not included in our real data comparison. 

\begin{table*}[htbp]
\centering
\caption{Average permutation variable importance ranks of all baseline covariates in the INSULIN trial data obtained from the two LRF variants. Smaller ranks indicate higher importance.}
\label{tab:vimp_rank}
\vspace{0.5cm}
\begin{tabular}{lcc}
\toprule
\textbf{Baseline Covariate} & \textbf{LRF--PACE--2} & \textbf{LRF--adaptiveLMM} \\
\midrule
Baseline A1C                           & 1  & 1  \\
Fasting Blood Glucose                  & 2  & 6  \\
Morning Fasting Glucose                & 3  & 7  \\
Noon 2-h Post-Lunch Glucose            & 4  & 4  \\
Noon Fasting Glucose                   & 5  & 3  \\
Evening Fasting Glucose                & 6  & 2  \\
Nighttime 3 a.m. Glucose               & 7  & 9  \\
Evening 2-h Post-Dinner Glucose        & 8  & 5  \\
Morning 2-h Post-Breakfast Glucose     & 9  & 8  \\
BMI                                    & 10 & 11  \\
Height                                 & 11 & 16  \\
Weight                                 & 12 & 10  \\
Systolic Blood Pressure                & 13 & 12  \\
Heart Rate                             & 14 & 19  \\
Duration of Diabetes                   & 15 & 18  \\
Therapy                                & 16 & 17  \\
Adiponectin                            & 17 & 14  \\
Diastolic Blood Pressure               & 18 & 15  \\
Fasting Insulin                        & 19 & 13  \\
\bottomrule
\end{tabular}
\end{table*}

Table~\ref{tab:vimp_rank} summarizes the average variable importance ranks across 10 cross-validation runs obtained from the two LRF variants. Although the exact ranks vary slightly between the two methods, both variants reveal a highly consistent group structure. First, baseline HbA1c is identified as the most influential covariate, highlighting its central role in influencing HbA1c response durability in type 2 diabetes. This finding is consistent with numerous previous studies that differed substantially in their analytical approaches, treatment regimens, scientific questions, and study populations, yet consistently reported greater treatment-induced reductions in HbA1c among patients with higher baseline HbA1c levels \citep{sato2024sglt2, esposito2014b, yin2016sitagliptin, esposito2015nomogram}. Second, glucose-related measurements, including fasting blood glucose and the SMBG profile (ranks 2--9), form the next most important group. This finding further validates previous studies showing that fasting and postprandial glucose levels are strongly associated with treatment-related HbA1c responses \citep{woerle2004diagnostic, vittal2021comparative, liao2021fasting, chandra2025explainable}. Third, anthropometric variables, including BMI, weight, and height, exhibit moderate importance. This result aligns with previous studies showing associations between obesity-related measures, insulin sensitivity, and glycemic control in patients with T2D \citep{deng2025bmi, balkau2015credit, huh2014bmi}. Finally, hemodynamic and additional clinical variables, including blood pressure, heart rate, adiponectin, fasting insulin, and diabetes duration, consistently appear near the bottom of the rankings, suggesting relatively limited contributions to variation in HbA1c durability. The treatment assignment (i.e., Therapy) ranks only 16th and 17th under LRF--PACE--2 and LRF--adaptiveLMM, respectively. This finding provides additional support for the original INSULIN trial analysis, which reported only a slight mean treatment difference of approximately 0.1\% in HbA1c reduction between the two treatments \citep{buse2009durable}.

Beyond corroborating the broader diabetes literature, the proposed LRF framework uncovers novel insights that have not been reported in previous analyses of the INSULIN trial. Specifically, existing analyses based on scalar endpoint values often assigned substantially different importance ranks to covariates within the same clinical category, making the results difficult to interpret. In contrast, without incorporating any prior knowledge of clinical categories, the proposed LRF methods naturally organize covariates into clinically meaningful importance groups. Table~\ref{tab:vimp_rank} shows that baseline HbA1c emerges as the dominant factor, followed by the glucose-related, anthropometric, and hemodynamic categories. This category-level importance pattern highlights the clinical group effects influencing long-term HbA1c durability more clearly than inconsistent individual ranks. For example, \citet{wang2018learning} identified baseline HbA1c, BMI, and fasting blood glucose as highly influential variables, whereas individual SMBG measurements, weight, and height received substantially lower ranks. Similarly, \citet{doubleday2018algorithm} ranked baseline body weight and mid-morning glucose highly, whereas other SMBG measurements, BMI, and height were assigned relatively low importance. By contrast, the proposed LRF methods consistently assign high importance to the entire glucose-related and anthropometric categories, revealing a more coherent clinical structure. To avoid the potential confounding of multicollinearity among variables within the same category, we experiment the process using different values of \textit{mtry} (\textit{mtry} = 3 and 5). This category-level importance pattern remains highly stable.

\begin{table}[htbp]
\centering
\caption{Top-ranked two-way, three-way, and four-way interactions identified by the frequency-based interaction detection procedure obtained from the two LRF variants.}
\label{tab:lrf_interactions}
\vspace{0.5cm}
\begin{tabular}{ll}
\toprule
\textbf{LRF--PACE--2} & \textbf{LRF--adaptiveLMM} \\
\midrule
A1C $\times$ BMI
  & A1C $\times$ EFG \\

A1C $\times$ WT
  & A1C $\times$ E2hG \\

A1C $\times$ BMI $\times$ SBP
  & A1C $\times$ SBP $\times$ EFG \\

A1C $\times$ BMI $\times$ HT
  & A1C $\times$ SBP $\times$ M2hG \\

MFG $\times$ M2hG $\times$ E2hG $\times$ N3FG
  & FG $\times$ SBP $\times$ M2hG $\times$ E2hG \\

FG $\times$ MFG $\times$ NFG $\times$ EFG
  & BMI $\times$ M2hG $\times$ EFG $\times$ E2hG \\
\bottomrule
\end{tabular}

\vspace{0.5em}

\begin{minipage}{0.95\linewidth}
\footnotesize
\raggedright

\textit{Abbreviations:}

Metabolic covariates: A1C = baseline HbA1c; FG = fasting glucose.

SMBG measurements: MFG = morning fasting glucose; M2hG = morning 2-hour postprandial glucose; NFG = noon fasting glucose; EFG = evening fasting glucose; E2hG = evening 2-hour postprandial glucose; N3FG = 3 a.m.\ fasting glucose.

Anthropometric covariates: BMI = body mass index; WT = weight; HT = height.

Hemodynamic covariate: SBP = systolic blood pressure.
\end{minipage}

\end{table}

\begin{table}[htbp]
\centering
\caption{Average prediction and forecast errors in the INSULIN trial real data. Smaller values indicate higher predictive accuracy. Top Panel: prediction of the entire HbA1c trajectories over Weeks 1--24 for new subjects. Bottom Panel: forecast of future HbA1c trajectories over Weeks 11--14 based on observations over Weeks 1--10 for existing subjects.}
\label{tab:pred_error}
\vspace{0.5cm}

\resizebox{\columnwidth}{!}{%
\begin{tabular}{lccccc}
\toprule
\textbf{Task} & \textbf{SPLF} & \textbf{LongCART} & \textbf{LRF--adaptiveLMM} & \textbf{LRF--PACE} & \textbf{LRF--PACE--2} \\
\midrule
New-subject & 7.636 & 1.616 & 1.588 & -- & 1.553 \\
Future-time & 7.660 & 3.263 & 0.678 & 0.536 & -- \\
\bottomrule
\end{tabular}%
}

\end{table}

\begin{figure*}[!t]
\centering
\includegraphics[width=\linewidth]{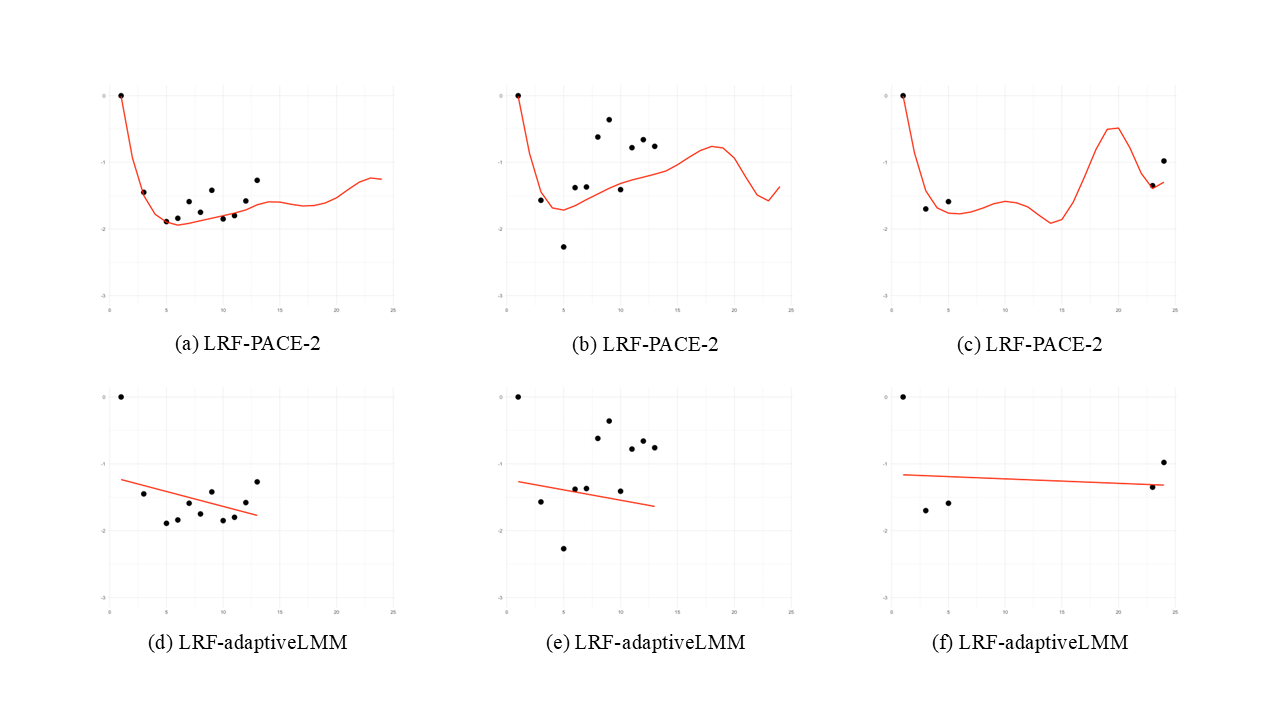}
\caption{Predicted trajectories for three randomly selected test subjects from the INSULIN trial data. For comparison purposes, each column displays the same subject, with LRF--PACE--2 in the top Panel and LRF--adaptiveLMM in the bottom Panel. The black points represent the original sparse and irregularly observed HbA1c measurements, and the red curves denote the corresponding predicted trajectories over Weeks~1--24.}
\label{fig:new-subject-prediction}
\end{figure*}

\begin{figure*}[!t]
\centering
\includegraphics[width=\linewidth]{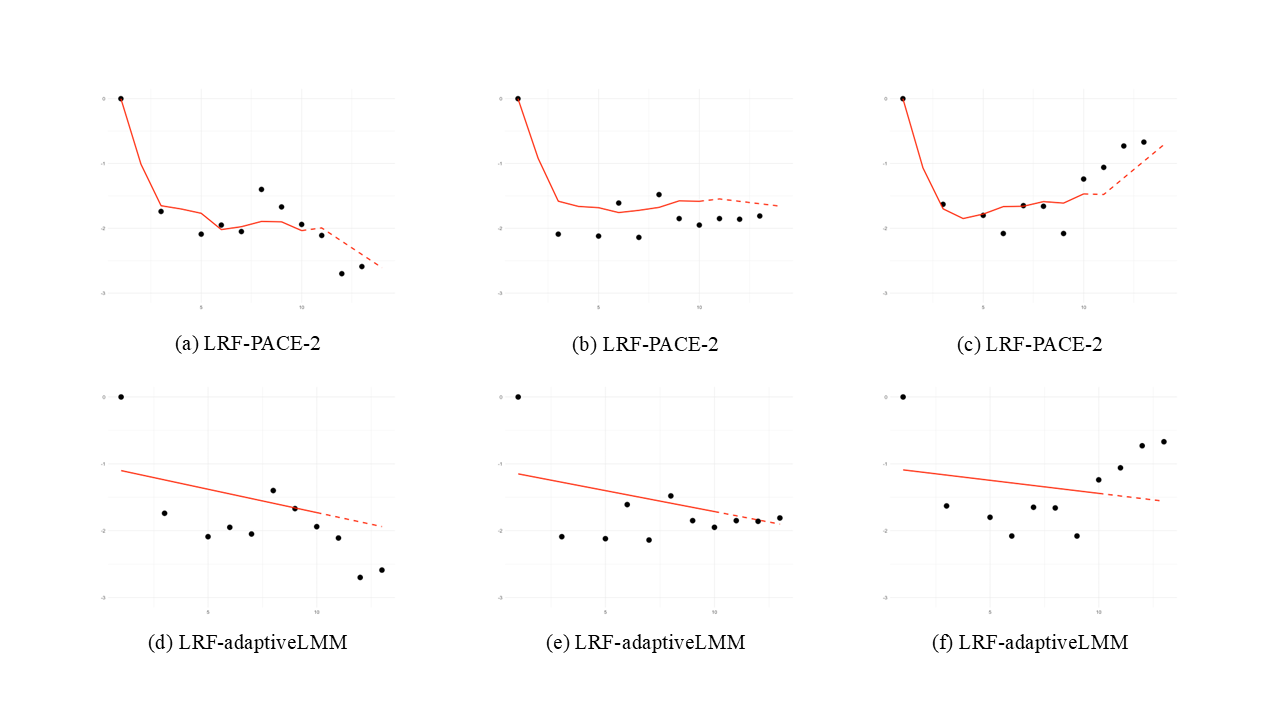}
\caption{Forecasted trajectories for three randomly selected training subjects from the INSULIN trial data. For comparison purpose, each column displays the same subject, with LRF--PACE--2 in the top Panel and LRF--adaptiveLMM in the bottom Panel. The black points represent the original sparse and irregularly observed HbA1c measurements. Solid red curves denote the estimated trajectories over the observed period (Weeks~1--10), whereas dashed red curves represent the forecasted future trajectories over Weeks~11--14.}
\label{fig:future-time-forecasting}
\end{figure*}

The proposed LRF framework further reveals another novel scientific finding through the frequency-based interaction procedure described in Section~\ref{sec:interaction_exp}. To the best of our knowledge, neither the general diabetes literature nor previous analyses of the INSULIN trial have reported comparable interaction effects on HbA1c responses to treatments. Table \ref{tab:lrf_interactions} summarizes the top ranked frequencies for two-way, three-way, and four-way interaction subsets that are obtained from the two LRF variants. Focusing on LRF--PACE--2, the most prominent two-way interactions involve baseline A1C paired with anthropometric category, and the leading three-way interactions further incorporate systolic blood pressure. In particular, the frequent co-occurrence of baseline A1C with anthropometric category suggests that its importance arises not only from its individual effect but also from its interactions with other covariates. The repeated appearance of systolic blood pressure in the top interaction sets, despite its relatively low individual importance rankings (13th in LRF--PACE--2 and 12th in LRF--adaptiveLMM), indicates that its contribution to HbA1c trajectories is driven primarily by interaction effects. The four-way interactions mainly involve glucose measures collected at different times throughout the day, which further explain their joint effects as a group. In addition, LRF--adaptiveLMM identifies other interactions involving baseline A1C, SMBG measures, and systolic blood pressure.

The top Panel of Table~\ref{tab:pred_error} summarizes the new-subject prediction performance across competing methods on the test data. Among all approaches, LRF--PACE--2 attains the smallest cross-validation prediction error ($1.553$), followed closely by LRF--adaptiveLMM ($1.588$), while SPLF performs substantially worse under this prediction setting ($7.636$). Figure~\ref{fig:new-subject-prediction} illustrates the predicted HbA1c trajectories for three randomly selected test subjects. For each subject, the trajectory predicted by LRF--PACE--2 closely follows the underlying dynamic trend reflected in the original discretely observed measurements. In Panels (a) and (b), the predicted trajectories capture the rapid decline in HbA1c immediately after treatment initiation followed by a slower increasing trend later in the study period. In Panel (c), the subject has only five observed measurements located near the beginning and the end of the study period. LRF--PACE--2 is still able to recover the early rapid decrease followed by an upward trend at the end. Note that the predicted trajectories do not necessarily closely follow all of the original observations because they are generated for test subjects using the learned covariate--trajectory relationships from the entire forest structure, rather than by simply smoothing each individual's observed longitudinal measurements. In comparison, LRF--adaptiveLMM produces trajectory estimates based on linear restriction and can only capture the overall decreasing trend of HbA1c after treatment initiation (Panels (d)--(f)). 

The bottom Panel of Table~\ref{tab:pred_error} reports the future-time forecasting performance across competing methods on the training data. Because the INSULIN dataset contains an almost empty interval between Weeks 15 and 22 (see Section~\ref{sec:data}), for evaluation purposes, we fit each model on Weeks 1--10 and forecast Weeks 11--14. Among all approaches, LRF--PACE attains the smallest average forecasting error ($0.536$) across 10 cross-validation replications, followed by LRF--adaptiveLMM ($0.678$), while SPLF yields the largest forecasting error ($7.66$). Figure~\ref{fig:future-time-forecasting} illustrates the forecasting results for three randomly selected training subjects. The forecasts produced by LRF--PACE at Weeks 11--14 closely align with those "future" observed values. For example, the forecast trajectories may exhibit a decreasing trend (Panel (a)), remain relatively flat (Panel (b)), or show a gradual increase (Panel (c)). In contrast, LRF--adaptiveLMM produces linear forecasts, with an overall linear decreasing trend across Panels (d), (e), and (f).

As an additional assessment of prediction accuracy, we compute Spearman's rank correlation coefficient $\rho$. At each evaluation time point, subjects are ranked separately according to their observed and predicted response values, and $\rho$ is calculated as the correlation between the two sets of ranks. Values of $\rho$ close to $1$ indicate strong agreement in the ordering between predicted and observed responses, whereas values near $0$ indicate little or no agreement. The reported $p$-value corresponds to the hypothesis test $H_0:\rho = 0$ versus $H_1:\rho \neq 0$.

\begin{table}[!t]
\centering
\caption{Average Spearman rank correlation coefficients ($\rho$) and corresponding $p$-values across 10 cross-validation replications for the INSULIN trial real data. Top Panel: prediction performance for new subjects. Bottom Panel: forecasting performance for existing subjects. Both Panels demonstrate only Week 12.}
\label{tab:rank_corr}
\vspace{0.5cm}
\setlength{\tabcolsep}{2pt}

\resizebox{\textwidth}{!}{%
\begin{tabular}{lccccc}
\toprule
\textbf{Task} & \textbf{SPLF} & \textbf{LongCART} & \textbf{LRF--adaptiveLMM} & \textbf{LRF--PACE} & \textbf{LRF--PACE--2} \\
\midrule
New-subject
& 0.35 ($p<0.01$)
& 0.46 ($p<0.01$)
& 0.26 ($p<0.01$)
& --
& 0.58 ($p<10^{-8}$) \\

Future-time
& 0.43 ($p<10^{-5}$)
& 0.17 ($p=0.11$)
& 0.45 ($p<10^{-6}$)
& 0.79 ($p<10^{-12}$)
& -- \\
\bottomrule
\end{tabular}%
}

\end{table}

\begin{figure*}[!t]
\centering
\includegraphics[width=0.9\textwidth]{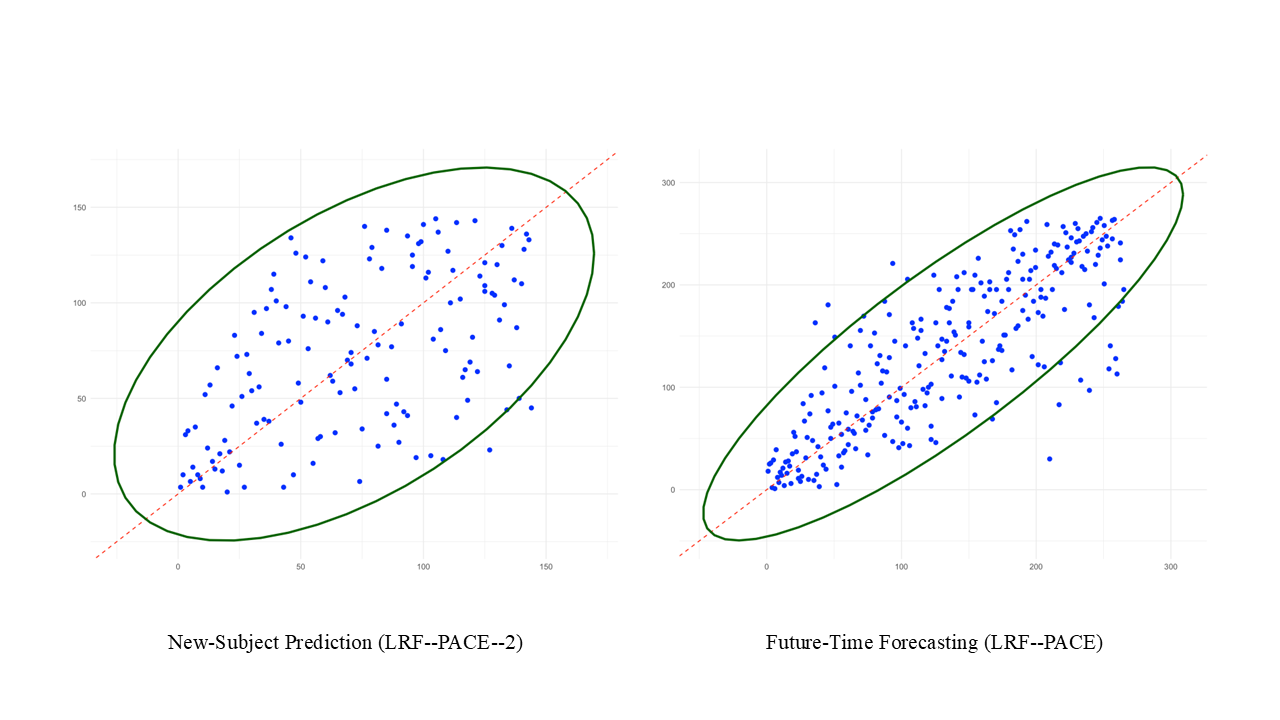}
\caption{Scatterplots of the ranks of the predicted/forecasted HbA1c response values versus the ranks of the corresponding originally observed response values from the INSULIN data for one cross-validation replication. The left Panel presents the predicted response values for test data using LRF--PACE--2, yielding a Spearman rank correlation of $\rho = 0.51$ ($p = 5\times10^{-11}$). The right Panel presents the forecasted response values for training data using LRF--PACE, yielding $\rho = 0.81$ ($p < 2.2\times10^{-16}$). The dashed red line indicates perfect agreement, and the green ellipse encloses 95\% of the observations. Both Panels demonstrate only Week~12.
}
\label{fig:lrf_pace_week12_rank}
\end{figure*}

Table~\ref{tab:rank_corr} summarizes the average Spearman rank correlations and corresponding $p$-values across 10 cross-validation runs at Week 12 for both prediction settings. Similar results are obtained at the other three weeks. 
For new-subject prediction, LRF--PACE--2 achieves the highest rank concordance with $\rho = 0.58$ ($p < 10^{-8}$). LongCART and SPLF attain moderate correlations of $0.46$ and $0.35$, respectively, while LRF--adaptiveLMM yields a weaker correlation of $0.26$, although still statistically significant ($p < 0.01$). For future-time forecasting, LRF--PACE again achieves the highest rank correlation with $\rho = 0.79$ ($p < 10^{-12}$). LRF--adaptiveLMM and SPLF obtain moderate correlations of $0.45$ and $0.43$, respectively, whereas LongCART shows weak rank agreement with $\rho = 0.17$ and a non-significant $p$-value ($p = 0.11$).

Figure~\ref{fig:lrf_pace_week12_rank} further illustrates the rank correlation results at Week~12. The left Panel presents the new-subject prediction results from one cross-validation replication obtained by LRF--PACE--2. The model is fitted using all available training observations from Weeks~1--24, and predictions are generated based on the baseline covariates of all test subjects, yielding a Spearman rank correlation of $\rho = 0.51$ ($p = 5\times10^{-11}$). The right Panel presents the future-time forecasting results from one cross-validation replication using LRF--PACE. The model is fitted on observations at Weeks~1--10 from the training subjects, and the forecast at Week~12, $\widetilde{\mathbf{Y}}_i(t_{\mathrm{new}})$, is obtained according to \eqref{eq:lrf_tree_pred_future} using the subject's baseline covariates, yielding a correlation of $\rho = 0.81$ ($p < 2.2\times10^{-16}$). Both correlations are highly statistically significant, with extremely small $p$-values, indicating strong agreement between the predicted and observed response rankings. The difference between the two correlation scores reflects the relative difficulty of the two tasks. Future-time forecasting only requires predicting outcomes over Weeks 11--14 for existing training subjects, whose observations from Weeks 1--10 have already been used to train the model. However, new-subject prediction requires predicting entire trajectories over Weeks~1--24 for previously unseen subjects using baseline covariates only, making it substantially more challenging than that of future-time forecasting.

\section{Discussion}
\label{sec:discussion}

In this article, we propose a novel Longitudinal Random Forest framework that jointly models sparse, irregular, and heterogeneous longitudinal response trajectories and their associations with scalar covariates within a unified framework, while flexibly accommodating complex nonlinear and interactive covariate effects. We also develop a trajectory-based splitting criterion that selects candidate splits using the size-weighted integrated squared difference between the representative trajectories of the resulting child nodes. To enable more comprehensive interpretation, we introduce a frequency-based interaction exploration strategy that efficiently identifies covariate interactions of arbitrary finite order within a single procedure.

The LRF framework includes two variants that differ in their node-wise trajectory estimation strategies: (i) LRF--PACE and (ii) LRF--adaptiveLMM. This design gives researchers the flexibility to select the node-wise smoother that best matches their data characteristics and analysis preferences. LRF--PACE adopts a fully nonparametric smoothing strategy, in which covariate information is incorporated indirectly through the tree structure. In contrast, LRF--adaptiveLMM is a semiparametric approach that explicitly incorporates the covariates that are selected along the path from the root to the corresponding node into the fixed-effects term of a linear mixed-effects model at each node. The node-specific model structure is then combined with subject-specific random effects to enable individualized trajectory estimation for each subject at each node. Compared with LRF--PACE, LRF--adaptiveLMM imposes a strict linear trajectory structure and is therefore less flexible; however, it may be preferred when researchers wish to retain the interpretability and convenience of the popular mixed-effects models while leveraging advances in machine learning.

Overall, LRF--PACE (or LRF--PACE--2) achieves the strongest performance in variable selection, entire trajectory prediction, and future-time forecast among all competing methods across a wide range of simulation settings, including scenarios with severe sparsity. The real-data application to the longitudinal INSULIN trial dataset further demonstrates the ability of the proposed LRF framework to uncover novel findings on treatment progression and long-term durability that are not identified in previous analyses of the same dataset using scalar endpoints. The proposed LRF framework has important implications for personalized medicine, where treatment decisions rely on individualized patient characteristics rather than uniform treatment strategies. In addition to the 19 baseline covariates analyzed in this study, the proposed LRF framework has a broad application scope and can readily incorporate a much larger range of patient characteristics, including genetic, lifestyle, diet, environmental, and other factors. This flexibility enhances its potential to support personalized medicine by providing a more comprehensive understanding of treatment durability.

\section{Conflicts of Interest}

The authors declare no conflicts of interest.

\bibliographystyle{plainnat}
\bibliography{LRF_bib}

@article{fan2007semiparametric,
  author    = {Fan, Jianqing and Huang, Tao and Li, Runze},
  title     = {Analysis of Longitudinal Data With Semiparametric Estimation of Covariance Function},
  journal   = {Journal of the American Statistical Association},
  year      = {2007},
  volume    = {102},
  number    = {478},
  pages     = {632--641},
  doi       = {10.1198/016214507000000095}
}

@article{DieuBriganti2025GEE,
  title   = {Biostatistics of generalized estimating equations in developmental medicine and child neurology},
  author  = {Dieu, Camille Eug{\'e}nie and Briganti, Giovanni},
  journal = {Developmental Medicine \& Child Neurology},
  year    = {2026},
  volume  = {68},
  number  = {6},
  pages   = {767--773},
  doi     = {10.1111/dmcn.70060},
  pmid    = {41204445}
}

@article{Frias2021Tirzepatide,
  author    = {Fr{\'\i}as, Juan P. and Davies, Melanie J. and Rosenstock, Julio and others},
  title     = {Tirzepatide versus semaglutide once weekly in patients with type 2 diabetes},
  journal   = {New England Journal of Medicine},
  year      = {2021},
  volume    = {385},
  number    = {6},
  pages     = {503--515},
  doi       = {10.1056/NEJMoa2107519}
}

@article{Sorli2017SUSTAIN3,
  author    = {Sorli, Carolina and Harashima, Shin and Tsoukas, Marek and Unger, Jochen and Kadowaki, Takashi and Araki, Eiji and Trautmann, Martin and Klimontov, Vadim V. and Kalra, Sanjay and Peterson, Suzanne and others},
  title     = {Efficacy and safety of once-weekly semaglutide vs exenatide ER in type 2 diabetes (SUSTAIN 3)},
  journal   = {Lancet Diabetes \& Endocrinology},
  year      = {2017},
  volume    = {5},
  number    = {4},
  pages     = {251--260},
  doi       = {10.1016/S2213-8587(17)30013-X}
}

@article{Dungan2014AWARD6,
  author    = {Dungan, Kathleen M. and Povedano, Silvio Torro and Forst, Thomas and others},
  title     = {Once-weekly dulaglutide versus once-daily liraglutide in type 2 diabetes (AWARD-6)},
  journal   = {Lancet},
  year      = {2014},
  volume    = {384},
  number    = {9951},
  pages     = {1349--1357},
  doi       = {10.1016/S0140-6736(14)60976-4}
}

@article{McCoy2023Trajectories,
  author    = {McCoy, Renee G. and Faust, Lisa and Heien, Heather C. and Patel, Shilpa and Caffo, Brian and Ngufor, Che},
  title     = {Longitudinal trajectories of glycemic control among U.S. adults with newly diagnosed diabetes},
  journal   = {Diabetes Research and Clinical Practice},
  year      = {2023},
  volume    = {205},
  pages     = {110989},
  doi       = {10.1016/j.diabres.2023.110989},
  pmid      = {37918637},
  pmcid     = {PMC10842883}
}

@article{sato2024sglt2,
  title        = {Model-based meta-analysis of HbA1c reduction across SGLT2 inhibitors using dose adjusted by urinary glucose excretion},
  author       = {Sato, Hiroshi and Ishikawa, Akira and Yoshioka, Hiroyuki and others},
  journal      = {Scientific Reports},
  volume       = {14},
  pages        = {24695},
  year         = {2024},
  publisher    = {Nature Publishing Group},
  doi          = {10.1038/s41598-024-76256-6},
  url          = {https://doi.org/10.1038/s41598-024-76256-6}
}

@article{esposito2014b,
  author  = {Esposito, Katherine and Chiodini, Paolo and Capuano, Annalisa and Maiorino, Maria Ida and Bellastella, Giuseppe and Giugliano, Dario},
  year    = {2014},
  title   = {Baseline glycemic parameters predict the hemoglobin {A1c} response to {DPP-4} inhibitors: meta-regression analysis of 78 randomized controlled trials with 20,053 patients},
  journal = {Endocrine},
  volume  = {46},
  number  = {1},
  pages   = {43--51},
  doi     = {10.1007/s12020-013-0090-0}
}

@article{yin2016sitagliptin,
  title   = {Quantitative Models for Evaluating the Correlation between Baseline {HbA1c} Levels and Sitagliptin as Monotherapy or Dual Therapy Treatment in Type 2 Diabetes: A Meta-Regression Analysis},
  author  = {Yin, Jiajing and Lin, Yi and Gu, Mingyu and Peng, Yongde},
  journal = {International Journal of Diabetes and Clinical Research},
  volume  = {3},
  number  = {3},
  pages   = {051},
  year    = {2016},
  doi     = {10.23937/2377-3634/1410051}
}

@article{esposito2015nomogram,
  title        = {Nomogram to estimate the HbA1c response to different DPP-4 inhibitors in type 2 diabetes: a systematic review and meta-analysis of 98 trials with 24,163 patients},
  author       = {Esposito, Katherine and others},
  journal      = {BMJ Open},
  volume       = {5},
  number       = {2},
  pages        = {e005892},
  year         = {2015},
  doi          = {10.1136/bmjopen-2014-005892},
  url          = {https://bmjopen.bmj.com/content/5/2/e005892}
}

@article{vittal2021comparative,
  title        = {A Comparative Study of Correlation of Random, Fasting and Postprandial Blood Glucose with Glycated Haemoglobin: A Multicentre Study},
  author       = {Vittal, B. G. and Patil, Mahantesh and Abhijith, D.},
  journal      = {Journal of Medical Science and Health},
  year         = {2021},
  volume       = {7},
  number       = {3},
  pages        = {26--31}
}

@article{liao2021fasting,
  title   = {Fasting and postprandial plasma glucose contribution to glycated haemoglobin and time in range in people with type 2 diabetes on basal and bolus insulin therapy: Results from a pooled analysis of insulin lispro clinical trials},
  author  = {Liao, Birong and Chen, Yun and Chigutsa, Farai and Piras de Oliveira, Carolina},
  journal = {Diabetes, Obesity and Metabolism},
  year    = {2021},
  volume  = {23},
  number  = {7},
  pages   = {1571--1579},
  doi     = {10.1111/dom.14370},
  pmid    = {33687790},
  pmcid   = {PMC8252747}
}

@article{woerle2004diagnostic,
  title        = {Diagnostic and therapeutic implications of relationships between fasting, 2-hour postchallenge plasma glucose and hemoglobin A1c values},
  author       = {Woerle, Hans J. and Pimenta, Walter P. and Meyer, Christian and others},
  journal      = {Archives of Internal Medicine},
  volume       = {164},
  number       = {15},
  pages        = {1627--1632},
  year         = {2004},
  doi          = {10.1001/archinte.164.15.1627}
}

@article{chandra2025explainable,
  title        = {Explainable prediction of long-term glycated hemoglobin response change in Finnish patients with type 2 diabetes following drug initiation using evidence-based machine learning approaches},
  author       = {Chandra, Gaurav and Lavikainen, Petteri and Siirtola, Pekka and Tamminen, Sanna and Ihalapathirana, Anuradha and Laatikainen, Tiina and R{\"o}ning, Juha},
  journal      = {Clinical Epidemiology},
  volume       = {17},
  pages        = {225--240},
  year         = {2025},
  doi          = {10.2147/CLEP.S505966},
  url          = {https://doi.org/10.2147/CLEP.S505966}
}

@article{deng2025bmi,
  title        = {Association Between Body Mass Index and Glycemic Control in Type 2 Diabetes Mellitus: A Cross-Sectional Study},
  author       = {Deng, L. and Jia, L. and Wu, X. L. and Cheng, M.},
  journal      = {Diabetes, Metabolic Syndrome and Obesity: Targets and Therapy},
  volume       = {18},
  pages        = {555--563},
  year         = {2025},
  month        = feb,
  doi          = {10.2147/DMSO.S508365},
  pmid         = {40007519},
  pmcid        = {PMC11853989},
  url          = {https://doi.org/10.2147/DMSO.S508365}
}

@article{balkau2015credit,
  title        = {Predictors of HbA1c over 4 years in people with type 2 diabetes starting insulin therapies: The CREDIT study},
  author       = {Balkau, Beverley and Calvi-Gries, Fran\c{c}oise and Freemantle, Nick and Vincent, Maya and Pilorget, Val{\'e}rie and Home, Philip D.},
  journal      = {Diabetes Research and Clinical Practice},
  volume       = {108},
  number       = {3},
  pages        = {432--440},
  year         = {2015},
  doi          = {10.1016/j.diabres.2015.02.010},
  url          = {https://www.sciencedirect.com/science/article/pii/S0168822715001357}
}

@article{huh2014bmi,
  title        = {The relationship between BMI and glycated albumin to glycated hemoglobin (GA/A1c) ratio according to glucose tolerance status},
  author       = {Huh, Jae Hyun and Kim, Kyu Jeung and Lee, Byung-Wan and Kim, Dae-Won and Kang, Eun Seok and Cha, Bong Soo and Lee, Hyeong Cheol},
  journal      = {PLoS ONE},
  volume       = {9},
  number       = {2},
  pages        = {e89478},
  year         = {2014},
  doi          = {10.1371/journal.pone.0089478},
  pmid         = {24586809},
  pmcid        = {PMC3938490},
  url          = {https://doi.org/10.1371/journal.pone.0089478}
}

@article{Handley2025BMJOpen,
  author    = {Handley, Daniel and Gillett, Alexander C. and Bala, Rahul and Tyrrell, John and Lewis, Cathryn M.},
  title     = {Latent class growth mixture modeling of HbA1C trajectories identifies individuals at high risk of developing complications of type 2 diabetes mellitus in the {UK Biobank}},
  journal   = {BMJ Open Diabetes Research \& Care},
  year      = {2025},
  volume    = {13},
  number    = {5},
  pages     = {e004826},
  doi       = {10.1136/bmjdrc-2024-004826},
  pmid      = {40921490},
  pmcid     = {PMC12421182}
}

@article{savieri2025limmcov,
  author    = {Savieri, P. and Stas, L. and Barbé, K.},
  title     = {LiMMCov: An interactive research tool for efficiently selecting covariance structures in linear mixed models using insights from time series analysis},
  journal   = {PLOS ONE},
  year      = {2025},
  volume    = {20},
  number    = {6},
  pages     = {e0325834},
  doi       = {10.1371/journal.pone.0325834},
  pmid      = {40498721},
  pmcid     = {PMC12157095},
  publisher = {Public Library of Science}
}

@article{kundu2019regtree,
  title={Regression Trees for Longitudinal Data with Baseline Covariates},
  author={Kundu, Malay Ghosh and Harezlak, Jaroslaw},
  journal={Biostatistics \& Epidemiology},
  volume={3},
  number={1},
  pages={1--22},
  year={2019},
  doi={10.1080/24709360.2018.1557797},
  pmid={30693349},
  pmcid={PMC6347409}
}

@Manual{NeufeldHeggeseth2025,
  title        = {splinetree: Longitudinal Regression Trees and Forests},
  author       = {Anna Neufeld and Brianna Heggeseth},
  year         = {2025},
  note         = {R package version 0.2.0, https://CRAN.R-project.org/package=splinetree},
  url          = {https://CRAN.R-project.org/package=splinetree}
}

@article{doubleday2018algorithm,
  title        = {An Algorithm for Generating Individualized Treatment Decision Trees and Random Forests},
  author       = {Doubleday, Kevin and Zhou, Hua and Fu, Haoda and Zhou, Jin},
  journal      = {Journal of Computational and Graphical Statistics},
  volume       = {27},
  number       = {4},
  pages        = {849--860},
  year         = {2018},
  doi          = {10.1080/10618600.2018.1451337},
  note         = {PMCID: PMC7274035}
}

@article{YuLambert1999,
  title        = {Fitting Trees to Functional Data, with an Application to Time-of-Day Patterns},
  author       = {Yan Yu and Diane Lambert},
  journal      = {Journal of Computational and Graphical Statistics},
  volume       = {8},
  number       = {4},
  pages        = {749--762},
  year         = {1999},
  doi          = {10.1080/10618600.1999.10474847}
}

@article{hajjem2011mixed,
  title={Mixed effects regression trees for clustered data},
  author={Hajjem, A. and Bellavance, F. and Larocque, D.},
  journal={Statistics and Probability Letters},
  volume={81},
  number={4},
  pages={451--459},
  year={2011}
}

@article{hajjem2014mixed,
  title={Mixed effects random forest for clustered data},
  author={Hajjem, A. and Bellavance, F. and Larocque, D.},
  journal={Journal of Statistical Computation and Simulation},
  volume={84},
  number={6},
  pages={1313--1328},
  year={2014}
}

@article{buse2009durable,
    author = {Buse, John B. and Wolffenbuttel, Bruce H.R. and Herman, William H. and Shemonsky, Natalie K. and Jiang, Honghua H. and Fahrbach, Jessie L. and Scism-Bacon, Jamie L. and Martin, Sherry A.},
    title = {DURAbility of Basal Versus Lispro Mix 75/25 Insulin Efficacy (DURABLE) Trial 24-Week Results: Safety and efficacy of insulin lispro mix 75/25 versus insulin glargine added to oral antihyperglycemic drugs in patients with type 2 diabetes},
    journal = {Diabetes Care},
    volume = {32},
    number = {6},
    pages = {1007-1013},
    year = {2009},
    month = {03},
    issn = {0149-5992},
    doi = {10.2337/dc08-2117},
    url = {https://doi.org/10.2337/dc08-2117},
    eprint = {https://diabetesjournals.org/care/article-pdf/32/6/1007/603424/zdc00609001007.pdf},
}

@article{laird1982random,
  title        = {Random-effects models for longitudinal data},
  author       = {Laird, Nan M. and Ware, James H.},
  journal      = {Biometrics},
  volume       = {38},
  number       = {4},
  pages        = {963--974},
  year         = {1982},
  month        = dec,
  issn         = {0006-341X},
  pmid         = {7168798}
}

@article{buse2011durable,
  author  = {Buse, John B. and Wolffenbuttel, Bruce H. R. and Herman, William H. and Hippler, Stephen and Martin, Sherry A. and Jiang, Honghua H. and Shenouda, Sylvia K. and Fahrbach, Jessie L.},
  title   = {Durability of basal versus lispro mix 75/25 insulin efficacy (DURABLE) trial: comparing the durability of lispro mix 75/25 and glargine},
  journal = {Diabetes Care},
  volume  = {34},
  number  = {2},
  pages   = {249--255},
  year    = {2011},
  month   = {February},
  doi     = {10.2337/dc10-1701},
  pmid    = {21270100},
  pmcid   = {PMC3024370}
}

@article{wang2018learning,
  author  = {Wang, Lu and Fu, Haoda and Zeng, Donglin},
  title   = {Learning optimal personalized treatment rules in consideration of benefit and risk: with an application to treating type 2 diabetes patients with insulin therapies},
  journal = {Journal of the American Statistical Association},
  volume  = {113},
  number  = {521},
  pages   = {1--13},
  year    = {2018},
  doi     = {10.1080/01621459.2017.1330207}
}

@book{diggle2002analysis,
  title={Analysis of Longitudinal Data},
  author={Diggle, Peter J. and Heagerty, Patrick and Liang, Kung-Yee and Zeger, Scott L.},
  year={2002},
  publisher={Oxford University Press},
  edition={2}
}

@article{yao2005functional,
  title={Functional data analysis for sparse longitudinal data},
  author={Yao, Fang and M{\"u}ller, Hans-Georg and Wang, Jane-Ling},
  journal={Journal of the American Statistical Association},
  volume={100},
  number={470},
  pages={577--590},
  year={2005},
  publisher={Taylor \& Francis},
  doi={10.1198/016214504000001745}
}

@article{liang1986longitudinal,
  title={Longitudinal data analysis using generalized linear models},
  author={Liang, Kung-Yee and Zeger, Scott L.},
  journal={Biometrika},
  volume={73},
  number={1},
  pages={13--22},
  year={1986},
  publisher={Oxford University Press},
  doi={10.1093/biomet/73.1.13}
}

@book{fitzmaurice2011applied,
  title        = {Applied Longitudinal Analysis},
  author       = {Fitzmaurice, Garrett M. and Laird, Nan M. and Ware, James H.},
  year         = {2011},
  edition      = {2nd},
  publisher    = {John Wiley \& Sons},
  address      = {Hoboken, NJ},
  pages        = {752},
  isbn         = {978-0-470-38027-7}
}

@article{caruana2015longitudinal,
  title     = {Longitudinal studies},
  author    = {Caruana, E. J. and Roman, M. and Hern{\'a}ndez-S{\'a}nchez, J. and Solli, P.},
  journal   = {Journal of Thoracic Disease},
  year      = {2015},
  volume    = {7},
  number    = {11},
  pages     = {E537--E540},
  doi       = {10.3978/j.issn.2072-1439.2015.10.63},
  pmid      = {26716051},
  pmcid     = {PMC4669300}
}

@article{sun2022idf,
  title     = {IDF Diabetes Atlas: Global, regional and country-level diabetes prevalence estimates for 2021 and projections for 2045},
  author    = {Sun, H. and Saeedi, P. and Karuranga, S. and Pinkepank, M. and Ogurtsova, K. and Duncan, B. B. and Stein, C. and Basit, A. and Chan, J. C. N. and Mbanya, J. C. and Pavkov, M. E. and Ramachandaran, A. and Wild, S. H. and James, S. and Herman, W. H. and Zhang, P. and Bommer, C. and Kuo, S. and Boyko, E. J. and Magliano, D. J.},
  journal   = {Diabetes Research and Clinical Practice},
  volume    = {183},
  pages     = {109119},
  year      = {2022},
  doi       = {10.1016/j.diabres.2021.109119},
  pmid      = {34879977},
  pmcid     = {PMC11057359},
  note      = {Erratum in \emph{Diabetes Research and Clinical Practice}, 2023, 204:110945}
}

@article{sherwani2016significance,
  title     = {Significance of HbA1c Test in Diagnosis and Prognosis of Diabetic Patients},
  author    = {Sherwani, S. I. and Khan, H. A. and Ekhzaimy, A. and Masood, A. and Sakharkar, M. K.},
  journal   = {Biomarker Insights},
  volume    = {11},
  pages     = {95--104},
  year      = {2016},
  doi       = {10.4137/BMI.S38440},
  pmid      = {27398023},
  pmcid     = {PMC4933534}
}

@incollection{eyth2025hemoglobin,
  title        = {Hemoglobin A1C},
  author       = {Eyth, E. and Zubair, M. and Naik, R.},
  booktitle    = {StatPearls [Internet]},
  publisher    = {StatPearls Publishing},
  address      = {Treasure Island, FL},
  year         = {2025},
  note         = {[Updated 2025 Jun 2]. Available from: \url{https://www.ncbi.nlm.nih.gov/books/NBK549816/}}
}

@article{Tabarraei2024,
  author    = {Tabarraei, Yaser and Keshtkar, Abbas Ali and Yekaninejad, Mir Saeed and Rahimi, Najme and Dowlatabadi, Yousef and Azam, Kamal},
  title     = {A Longitudinal Examination of Blood Sugar Dynamics in Diabetes and Non-Diabetes Using Growth Curve Model: The Sabzevar Persian Cohort Study},
  journal   = {Advanced Biomedical Research},
  year      = {2024},
  volume    = {13},
  number    = {1},
  pages     = {30}
}

@article{Kim2019,
  author    = {Kim, K. and Unni, S. and Brixner, D. I. and Thomas, S. M. and Olsen, C. J. and Sterling, K. L. and Mitchell, M. and McAdam-Marx, C.},
  title     = {Longitudinal changes in glycated haemoglobin following treatment intensification after inadequate response to two oral antidiabetic agents in patients with type 2 diabetes},
  journal   = {Diabetes, Obesity and Metabolism},
  year      = {2019},
  volume    = {21},
  number    = {7},
  pages     = {1725--1733},
  doi       = {10.1111/dom.13694},
  pmid      = {30848039},
  pmcid     = {PMC6618330}
}

@article{mccarthy2017personalised,
  author  = {McCarthy, Mark I.},
  title   = {Painting a new picture of personalised medicine for diabetes},
  journal = {Diabetologia},
  year    = {2017},
  volume  = {60},
  number  = {5},
  pages   = {793--799},
  doi     = {10.1007/s00125-017-4210-x}
}

@article{pearson2019multifaceted,
  author  = {Pearson, Ewan R.},
  title   = {Type 2 diabetes: a multifaceted disease},
  journal = {Diabetologia},
  year    = {2019},
  volume  = {62},
  number  = {6},
  pages   = {1107--1112},
  doi     = {10.1007/s00125-019-4909-y}
}

@article{QIU2025102367,
title = {Phenotypic heterogeneity of type 2 diabetes and risks of all-cause and cause-specific mortality},
journal = {Cell Reports Medicine},
pages = {102367},
year = {2025},
issn = {2666-3791},
doi = {https://doi.org/10.1016/j.xcrm.2025.102367},
url = {https://www.sciencedirect.com/science/article/pii/S2666379125004409},
author = {Zixin Qiu and Frank Qian and Jun Liu and Rui Li and Hancheng Yu and Yue Wang and Xiao Zhang and Tingting Geng and Xuefeng Yu and Oscar H. Franco and An Pan and Maigeng Zhou and Kai Huang and Gang Liu}
}

@article{Tee2023trajectory,
  author    = {Tee, Clarence and Xu, Haiyan and Fu, Xiuju and Cui, Di and Jafar, Tazeen H. and Bee, Yong Mong},
  title     = {Longitudinal HbA1c trajectory modelling reveals the association of HbA1c and risk of hospitalization for heart failure for patients with type 2 diabetes mellitus},
  journal   = {PLOS ONE},
  year      = {2023},
  volume    = {18},
  number    = {1},
  pages     = {1--13},
  doi       = {10.1371/journal.pone.0275610},
  url       = {https://doi.org/10.1371/journal.pone.0275610},
  publisher = {Public Library of Science}
}

@article{yan2025comparison,
  title   = {Comparison of different approaches in handling missing data in longitudinal multiple-item patient-reported outcomes: a simulation study},
  author  = {Yan, Min and Zhou, Li and Zhao, Chao and Shi, Chen and Ou, Chao},
  journal = {Health and Quality of Life Outcomes},
  volume  = {23},
  number  = {1},
  pages   = {34},
  year    = {2025},
  doi     = {10.1186/s12955-025-02364-0}
}

@article{he2010missing,
  title   = {Missing data analysis using multiple imputation: getting to the heart of the matter},
  author  = {He, Yang},
  journal = {Circulation: Cardiovascular Quality and Outcomes},
  volume  = {3},
  number  = {1},
  pages   = {98--105},
  year    = {2010},
  doi     = {10.1161/CIRCOUTCOMES.109.875658}
}

@article{sassi-sayadi2026regulatory,
  author  = {Sassi-Sayadi, Mouna and Verweij, Pierre and Cornelisse, Peter},
  year    = {2026},
  title   = {Regulatory Experiences with the Use of Multiple Imputation for Missing Data in a Phase 3 Confirmatory Trial},
  journal = {Therapeutic Innovation \& Regulatory Science},
  volume  = {60},
  number  = {1},
  pages   = {8--14},
  doi     = {10.1007/s43441-025-00872-1}
}

@article{horton2007much,
  title   = {Much ado about nothing: {A} comparison of missing data methods and software to fit incomplete data regression models},
  author  = {Horton, Nicholas J. and Kleinman, Ken P.},
  journal = {The American Statistician},
  volume  = {61},
  number  = {1},
  pages   = {79--90},
  year    = {2007},
  doi     = {10.1198/000313007X172556}
}

@article{Aniley2019SemiParametric,
  author  = {Aniley, Tafere Tilahun and Debusho, Legesse Kassa and Nigusie, Zelalem Mehari and Yimer, Wondwosen Kassahun and Yimer, Belay Birlie},
  title   = {A semi‐parametric mixed models for longitudinally measured fasting blood sugar level of adult diabetic patients},
  journal = {BMC Medical Research Methodology},
  year    = {2019},
  volume  = {19},
  number  = {1},
  pages   = {13},
  doi     = {10.1186/s12874-018-0648-x}
}

@article{verbeke2014multivariate,
  author    = {Verbeke, Geert and Fieuws, Steffen and Molenberghs, Geert and Davidian, Marie},
  title     = {The analysis of multivariate longitudinal data: A review},
  journal   = {Statistical Methods in Medical Research},
  year      = {2014},
  volume    = {23},
  number    = {1},
  pages     = {42--59},
  doi       = {10.1177/0962280212445834},
  pmid      = {22523185},
  pmcid     = {PMC3404254}
}

@article{Mancl2001,
  author  = {Mancl, Lloyd A. and DeRouen, Timothy A.},
  title   = {A covariance estimator for GEE with improved small-sample properties},
  journal = {Biometrics},
  year    = {2001},
  volume  = {57},
  number  = {1},
  pages   = {126--134}
}

@article{wang2016covariance,
  author    = {Wang, M. and Kong, L. and Li, Z. and Zhang, L.},
  title     = {Covariance estimators for generalized estimating equations (GEE) in longitudinal analysis with small samples},
  journal   = {Statistics in Medicine},
  year      = {2016},
  volume    = {35},
  number    = {10},
  pages     = {1706--1721},
  doi       = {10.1002/sim.6817},
  pmid      = {26585756},
  pmcid     = {PMC4826860}
}

@article{vanderHorn2024,
  author    = {van der Horn, H. J. and Erhardt, E. B. and Dodd, A. B. and Nathaniel, U. and Wick, T. V. and McQuaid, J. R. and Ryman, S. G. and Vakhtin, A. A. and Meier, T. B. and Mayer, A. R.},
  title     = {A cautionary tale on the effects of different covariance structures in linear mixed effects modeling of fMRI data},
  journal   = {Human Brain Mapping},
  year      = {2024},
  volume    = {45},
  number    = {7},
  pages     = {e26699},
  doi       = {10.1002/hbm.26699},
  pmid      = {38726907},
  pmcid     = {PMC11082918}
}

@book{VerbekeMolenberghs2009,
  author    = {Verbeke, Geert and Molenberghs, Geert},
  title     = {Linear Mixed Models for Longitudinal Data},
  edition   = {2nd},
  publisher = {Springer},
  address   = {New York},
  year      = {2009},
  series    = {Springer Series in Statistics},
  doi       = {10.1007/978-0-387-77218-1}
}

@article{chen2015irregular,
  author    = {Chen, Yong and Ning, Jing and Cai, Chunyan},
  title     = {Regression analysis of longitudinal data with irregular and informative observation times},
  journal   = {Biostatistics},
  year      = {2015},
  volume    = {16},
  number    = {4},
  pages     = {727--739},
  doi       = {10.1093/biostatistics/kxv008},
  url       = {https://doi.org/10.1093/biostatistics/kxv008}
}

@article{Luo2018hba1c,
  author    = {Luo, Ming and Tan, Kenneth H. X. and Tan, Chuen Seng and Lim, Wei Yen and Tai, E. Shyong and Venkataraman, Kavita},
  title     = {Longitudinal trends in HbA1c patterns and association with outcomes: A systematic review},
  journal   = {Diabetes \& Metabolism Research and Reviews},
  year      = {2018},
  volume    = {34},
  number    = {6},
  pages     = {e3015},
  doi       = {10.1002/dmrr.3015},
  pmid      = {29663623},
  pmcid     = {PMC6175395}
}

@article{delporte2024analysing,
  author    = {Delporte, M. and Aerts, M. and Verbeke, G. and Molenberghs, G.},
  title     = {Analysing matched continuous longitudinal data: A review},
  journal   = {Statistical Methods in Medical Research},
  year      = {2024},
  volume    = {34},
  number    = {1},
  pages     = {170--179},
  doi       = {10.1177/09622802241300823},
  url       = {https://doi.org/10.1177/09622802241300823}
}

@article{lu2010covariance,
  author    = {Lu, Kaifeng and Mehrotra, Devan V.},
  title     = {Specification of covariance structure in longitudinal data analysis for randomized clinical trials},
  journal   = {Statistics in Medicine},
  year      = {2010},
  volume    = {29},
  number    = {4},
  pages     = {474--488},
  doi       = {10.1002/sim.3820},
  pmid      = {20020423}
}

@article{sela2012reem,
  title={RE-EM trees: A data mining approach for longitudinal and clustered data},
  author={Sela, Rebecca J. and Simonoff, Jeffrey S.},
  journal={Machine Learning},
  volume={86},
  number={2},
  pages={169--207},
  year={2012},
  publisher={Springer}
}

@manual{sela2021reemtree,
  title={REEMtree: Regression trees with random effects for longitudinal (panel) data},
  author={Sela, Rebecca J. and Simonoff, Jeffrey S. and Jing, Weixin},
  year={2021},
  note={R package version 0.90.4},
  url={https://cran.r-project.org/package=REEMtree}
}

@article{capitaine2021rf,
  title={Random forests for high-dimensional longitudinal data},
  author={Capitaine, Lo{\"i}c and Genuer, Robin and Thi{\'e}baut, Rodolphe},
  journal={Statistical Methods in Medical Research},
  volume={30},
  number={1},
  pages={166--184},
  year={2021},
  pmid={32772626},
  publisher={SAGE Publications}
}

@manual{capitaine2020longiturf,
  title={LongituRF: Random forests for longitudinal data},
  author={Capitaine, Lo{\"i}c},
  year={2020},
  note={R package version 0.9},
  url={https://cran.r-project.org/package=LongituRF}
}

@article{vickers2001analysing,
  title   = {Statistics notes: Analysing controlled trials with baseline and follow up measurements},
  author  = {Vickers, Andrew J. and Altman, Douglas G.},
  journal = {BMJ},
  volume  = {323},
  number  = {7321},
  pages   = {1123--1124},
  year    = {2001},
  doi     = {10.1136/bmj.323.7321.1123}
}

@article{mallinckrodt2008recommendations,
  title   = {Recommendations for the Primary Analysis of Continuous Endpoints in Longitudinal Clinical Trials},
  author  = {Mallinckrodt, Craig H. and Lane, Peter W. and Schnell, David and Peng, Yi and Mancuso, John P.},
  journal = {Therapeutic Innovation \& Regulatory Science},
  volume  = {42},
  number  = {4},
  pages   = {303--319},
  year    = {2008},
  doi     = {10.1177/009286150804200402}
}

@manual{FDA2023Covariates,
  author       = {{U.S. Food and Drug Administration}},
  title        = {Adjusting for Covariates in Randomized Clinical Trials for Drugs and Biological Products},
  year         = {2023},
  url          = {https://www.fda.gov/regulatory-information/search-fda-guidance-documents/adjusting-covariates-randomized-clinical-trials-drugs-and-biological-products}
}

@manual{EMA2015Covariates,
  author       = {{European Medicines Agency}},
  title        = {Guideline on Adjustment for Baseline Covariates in Clinical Trials},
  year         = {2015},
  url          = {https://www.ema.europa.eu/en/documents/scientific-guideline/guideline-adjustment-baseline-covariates-clinical-trials_en.pdf}
}

\end{document}